%% file: main.tex
\documentclass[conference]{IEEEtran}
\IEEEoverridecommandlockouts

\usepackage{cite}
\usepackage{amsmath,amssymb}
\usepackage{graphicx}
\usepackage{booktabs}
\usepackage{multirow}
\usepackage{xcolor}
\usepackage[hidelinks]{hyperref}

\begin{document}

\title{Space Over Statistic, Retraining Over Ranking: Auditing Training Labels Under Label-Only Backdoors}

\author{
\IEEEauthorblockN{Arnab Dey}
\IEEEauthorblockA{\textit{Department of Computer Science and Engineering} \\
\textit{Bangladesh University of Engineering and Technology}\\
Dhaka, Bangladesh \\
arnabdeykabya@gmail.com}
\and
\IEEEauthorblockN{Mohammad Mahfuzul Islam}
\IEEEauthorblockA{\textit{Department of Computer Science and Engineering} \\
\textit{Bangladesh University of Engineering and Technology}\\
Dhaka, Bangladesh \\
mahfuz@cse.buet.ac.bd}
}

\maketitle

\begin{abstract}
Some backdoor attacks never touch an image. FLIP only changes labels: on
CIFAR-10, flipping 3\% of them makes a ResNet-32 send 99.7\% of triggered
source-class images to the attacker's class. The usual defenses look for the
poison inside a model trained on those same labels, one already fitted to the
labels being checked. So when they struggle, is the problem the test they run,
or the space they run it in?

We answer with a controlled swap: we keep the test, the labels and the metric
fixed, change only the feature space, and judge each defense by retraining the
victim on what it keeps. The space decides. On dog$\rightarrow$cat the same
nearest-neighbor check scores AUROC 0.546 on the poisoned model's features,
0.967 on frozen DINOv2 features that never saw a label, and 0.995 on the same
network trained on clean labels, an oracle control rather than a defense. We
reproduce SPECTRE, the defense FLIP's authors recommend, to 90.94 / 0.97
against its published 91.47 / 1.7, then price it: on truck$\rightarrow$deer it
deletes 20\% of every class, while a prototype audit on the frozen features
deletes 6\% for higher clean accuracy and lower attack success. On three
harder pairs that audit leaves about a quarter of the attack standing. A count
read off encoder geometry and clean labels, before any attack is run,
separates the pair where the audit removes the attack from those where it only
partly does, though on a single zero-count pair.

Three results warn against the easy story. A better detection score does not
mean a safer model: on two of four pairs Confident Learning scores the higher
AUROC, and on dog$\rightarrow$cat leaves 2.6 times the attack success. The
prototype family, our best static performer, is the first an adaptive attacker
slips past. And past a point, catching more poisons stops reducing attack
success.

Check labels in a space that never learned them, and judge a defense by the
retrained model, not by its detection score.
\end{abstract}

\section{Introduction}
\label{sec:intro}

\begin{figure*}[t]
  \centering
  \includegraphics[width=\textwidth]{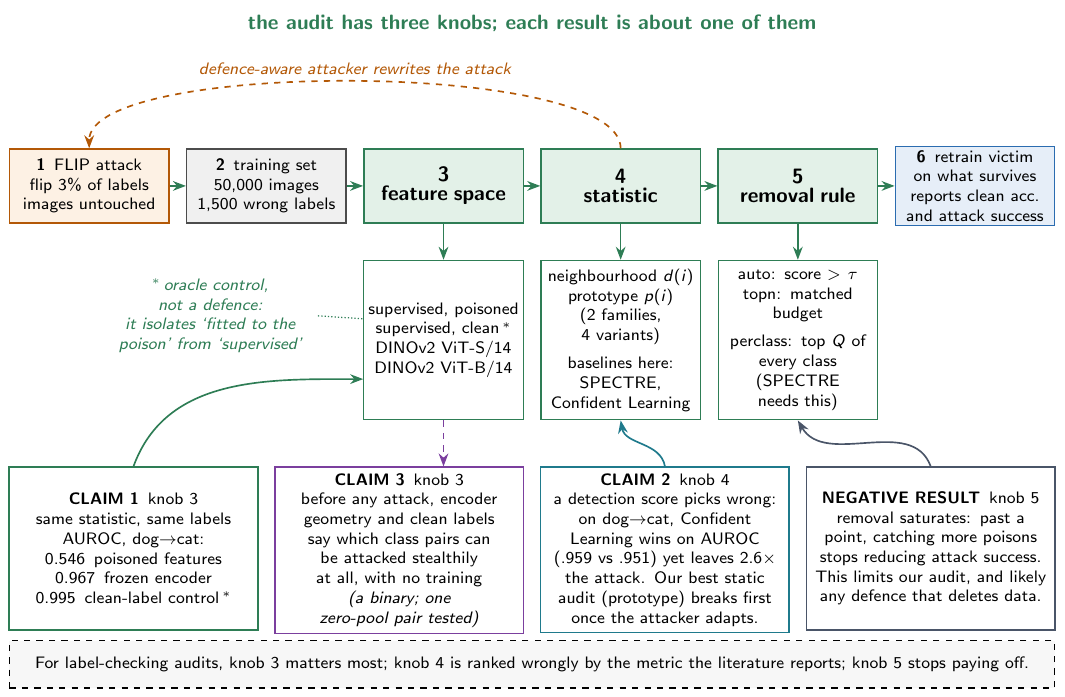}
  \caption{The pipeline, and where each claim acts. FLIP corrupts labels only,
  so the images reaching the defender are the ones that were collected. An
  audit then has three knobs: the space features live in, the statistic
  computed on them, and the rule turning scores into deletions. We turn one
  at a time, always retraining the victim rather than stopping at a ranking.
  The clean-label space ($^{*}$) is a control, not a defense: it holds the
  architecture fixed and changes only the labels the features were fitted
  to, separating ``supervised features are poor here'' from ``features fitted
  to the poison are poor here''. Two dashed arrows leave the main flow: the
  defense-aware attacker feeds back into the attack, and the stealthy-flip
  pool reads the encoder and clean labels before any attack exists. Each
  claim sits under the knob it concerns. The two statistic families at
  knob~4 are defined in \eqref{eq:knn} and \eqref{eq:proto}. Claim~1 is
  Section~\ref{sec:claim1}, Claim~2 and the negative result are
  Section~\ref{sec:claim2}, and Claim~3 is Section~\ref{sec:claim3}. AUROC
  values are \texttt{knn\_disagreement} on dog$\rightarrow$cat
  (Section~\ref{sec:claim1}).}
  \label{fig:overview}
\end{figure*}

Training labels often come from somewhere the model owner does not control.
Crowd workers annotate them, or a shared teacher model produces them, and
poisoning data at web scale is practical
today~\cite{carlini2024webscale,barreno2010security}. FLIP~\cite{jha2023flip}
shows how far an attacker gets with that access alone. It changes labels and
nothing else. That means every training image arrives exactly as it was collected
and on CIFAR-10 flipping 3\% of the labels installs a backdoor that fires on
99.7\% of triggered test images, while costing about three points of clean
accuracy. Unlike classical backdoor attacks~\cite{gu2017badnets}, there is no
tampered image for a defender to find. The only thing that is wrong is the
labels. Figure~\ref{fig:overview} lays out the whole setting. The attacker
changes labels; the training set reaches the defender with its images intact.
An audit then has three things a defender chooses: the space the features
live in, the statistic computed on that space, and the rule that turns scores
into deletions and this paper changes one of them at a time, always
retraining the victim rather than stopping at a detection score.

The defenses a practitioner would reach for all work the same way. Spectral
signatures, activation clustering, Deep $k$-NN and SPECTRE search for the
planted samples inside the representations of a model trained on the poisoned
data~\cite{tran2018spectral,chen2019activation,peri2020deepknn,hayase2021spectre}.
That is an awkward place to look, because the model was trained to fit the very
labels the defense is questioning. FLIP's own analysis says so directly: its
Fig.~9(c) shows the flipped pairs drifting into the target class in
representation space as training proceeds \cite{jha2023flip}. The defense is reading an instrument
the attack has bent.

It would be easy to conclude from this that nothing works, and that conclusion
would be wrong. FLIP's authors test three defenses on exactly this setting and
report SPECTRE at 91.47 clean accuracy with 1.7\% attack success (their
Table~16)~\cite{jha2023flip}, and we reproduce that result here. The useful question is therefore not whether a
defense exists. It is what makes one work, what it costs, and how a defender is
supposed to choose between the options. The trustworthiness question behind
all of this, and the one every experiment in this paper answers, is whether a
defender who holds no trusted data can decide which training labels to keep,
and what that decision costs in clean data destroyed and in the accuracy of
the model that is finally trained.

\paragraph{The gap} Four questions follow, and none has been answered.

First, the defenses above all differ in their statistic but read the same
kind of space~\cite{tran2018spectral,chen2019activation,peri2020deepknn,hayase2021spectre}.
So nobody knows which one is the bottleneck. Qi et al.\ show an adaptive
attacker can break that space on purpose~\cite{qi2023latent}. Nobody has
measured how weak it is with no attacker at all, holding the statistic
fixed, changing only the space, with a clean-label control to say whether
the problem is supervision or the poison.

Second, FLIP reports what SPECTRE protects~\cite{jha2023flip}, not what it
costs. SPECTRE sizes its cut from the poison count~\cite{hayase2021spectre},
which no defender has. What it deletes at a budget a defender can set is
unreported.

Third, audits are ranked by AUROC and AUPRC. A recent SoK calls these
metrics too narrow~\cite{abad2025sok}, and a benchmark shows the numbers
shift with setup~\cite{schwarzschild2021toxic}. Both argue across the
literature. Neither shows the ranking failing inside one attack, at equal
budget, on the retrained victim.

Fourth, a defender has two questions a score cannot answer: will this work
on my data, and does deleting more keep helping?

\paragraph{What we do} We audit labels against a frozen, public image encoder
that never sees a training label~\cite{oquab2024dinov2}, and we use that as an
instrument for taking the question apart rather than as a new algorithm. Our
contributions are:

\begin{itemize}
  \item \textbf{The space decides, not the statistic.} Holding the detector,
        the labels and the metric fixed and changing only the feature space,
        a neighborhood audit moves from AUROC 0.546 on the poisoned victim's
        own features to 0.967 on frozen features. The same architecture trained
        on \emph{clean} labels reaches 0.995, which locates the deficit in
        fitting the poison rather than in supervised learning
        (Section~\ref{sec:claim1}).

  \item \textbf{What the strongest published defense costs.} That SPECTRE stops
        FLIP is already known; what it costs to do so is not. We show it buys
        that protection by deleting 20\% of every class, while a
        label-independent audit reaches both a higher clean accuracy and a lower
        attack success after deleting 6\% which is a 3.4$\times$ smaller cut than SPECTRE. Our
        reimplementation matches the result FLIP report for it (90.94 / 0.97
        against 91.47 / 1.7)~\cite{jha2023flip}, which is what makes this a
        comparison against SPECTRE
        rather than against a weakened copy of it
        (Section~\ref{sec:spectre}).

  \item \textbf{Detection scores select the wrong audit.} We show this three
        ways: a better-ranking variant of our own statistic yields a worse
        model; Confident Learning~\cite{northcutt2021confident} attains a higher
        AUROC than our audit on two of four class pairs while leaving more than
        twice the attack success; and the audit that wins every static
        comparison is the one a defense-aware attacker dismantles first. The
        general finding is that detection quality is not a proxy for
        protection, so an audit chosen on detection scores alone can be the
        wrong one (Section~\ref{sec:claim2}).

  \item \textbf{A first step towards predicting where the audit applies.} A
        count computed from encoder geometry alone (no attack, no training)
        tells whether a stealthy attack on a given class pair is possible
        at all. On the one pair we tested where that count is zero, the audit
        removes the backdoor outright; on three pairs where it is not, about a
        quarter of the attack survives. We report this as a binary, and only
        as a binary: among the pairs where an attack exists the count carries
        no ordering, and the separation itself rests on a single zero-count
        pair (Section~\ref{sec:claim3}).

  \item \textbf{A ceiling on deletion-based defense.} Removal saturates: past
        a point, catching more poisons stops reducing attack success. This
        constrains our own method as much as any other that works by deleting
        data. Together with the finding that detection quality does not predict
        protection, this is the half of the paper we think is most useful.
\end{itemize}

We do not propose a new detection algorithm, and we do not claim robustness to
an adaptive attacker. Section~\ref{sec:limitations} states what the results do
not establish.

\section{Background and threat model}
\label{sec:background}

A backdoor attack leads to a model that is accurate on normal inputs, but predicts an attacker-chosen label when a trigger is present in the image. The standard recipe modifies both parts of a training example: the trigger is stamped onto a subset of images and they are relabelled to the target class~\cite{gu2017badnets,tran2018spectral}. FLIP~\cite{jha2023flip} wastes half of the image. Just changing the labels, it installs the same backdoor, and every training image reaches the victim exactly as it was collected. This is consistent with the usual way of creating labels: crowd-sourced annotation and knowledge distillation both give a third party control over labels but not over the data~\cite{barreno2010security}.

\subsection{How FLIP builds a label-only backdoor}

FLIP runs in three steps. It first trains \emph{expert} models on a
conventionally backdoored copy of the dataset (images carrying the trigger,
relabelled to the target) and records their training trajectories. It then
uses trajectory matching, borrowed from dataset
distillation~\cite{cazenavette2022trajectory}, to optimise soft labels
$\tilde{\ell}$ over the \emph{clean} images, so that a student trained on those
images and those labels follows the expert's trajectory. Finally it rounds the
soft labels into hard flips. Each training example $x_i$ with true label $y_i$
receives a score
\begin{equation}
  s(x_i) \;=\; \max_{c \neq y_i} \tilde{\ell}_{i,c} \;-\; \tilde{\ell}_{i,y_i},
  \label{eq:flip-score}
\end{equation}
the largest incorrect logit minus the logit of the correct class. The attacker
keeps the $m$ examples with the highest score and relabels each one to its own
$\arg\max_{c \neq y_i} \tilde{\ell}_{i,c}$. The victim then trains normally on
the clean images and the flipped labels.

Two properties of \eqref{eq:flip-score} are easy to overlook, and both limit
what a defense may assume. The score is computed over the whole training set
rather than inside one class, so nothing pushes the flips into the source
class. And the new label is whichever wrong class the experts happen to prefer,
not one fixed target. Concentration in a single source and target class is an
outcome of the optimisation, not a constraint. At $m = 1500$ we count 1297
flips landing on truck images and 1209 carrying the label deer for
truck$\rightarrow$deer; for dog$\rightarrow$cat the counts are 1416 and 1499. A
defense that assumes every poison sits in one class, or that all poisons share
one label, assumes something the attack does not promise.

\subsection{Metrics}

We keep the two metrics the attack paper uses. Clean test accuracy (CTA) is the
victim's accuracy on the untouched test set. Poison test accuracy (PTA) is the
fraction of triggered source-class test images the victim sends to the target
class, so it is the attack's success rate:
\begin{align}
  \mathrm{CTA} &= \frac{1}{|D_{\mathrm{test}}|}
     \sum_{(x,y) \in D_{\mathrm{test}}} \mathbf{1}\!\left[f(x) = y\right],
     \label{eq:cta}\\
  \mathrm{PTA} &= \frac{1}{|D_{\mathrm{src}}|}
     \sum_{x \in D_{\mathrm{src}}} \mathbf{1}\!\left[f(T(x)) = y_{\mathrm{tgt}}\right],
     \label{eq:pta}
\end{align}
where $f$ is the trained victim, $T$ applies the trigger, and $D_{\mathrm{src}}$
holds the 1\,000 test images of the source class. A defense wants CTA high and
PTA low.

PTA does not bottom out at zero. A model trained on entirely clean labels still
mixes up similar classes, so a few triggered source images reach the target
class through ordinary error. We call this the \emph{clean-label floor} and
measure it for each pair by training on clean labels: 0.001 on
truck$\rightarrow$deer and 0.061 on dog$\rightarrow$cat. PTA is not comparable
across pairs without it, because a clean model already confuses dogs with cats
and never confuses trucks with deer.

\subsection{Threat model}

\paragraph{Attacker} The attacker supplies labels and nothing else. Every image
is untouched. The budget is $m = 1500$ flips out of 50\,000 training examples,
or 3\%. The attacker knows the trigger and can afford to train expert models on
the training images. In our experiments the expert and the victim share the
ResNet-32 architecture~\cite{he2016resnet}; FLIP shows that this match is
not required~\cite{jha2023flip}.

\paragraph{Defense-aware attacker}  We also give the attacker the defender's statistic, the defender's encoder and the defender's own scoring code in Section~\ref{sec:claim2}, so that the two sides cannot silently diverge. This attacker is still limited. It decides \emph{which} examples to flip, not what to flip them to. It does not re-optimise the experts under the stealth constraint.
This is thus a lower bound on what a defense aware attacker may achieve.
In this paper, we report it as one and make no claims about its resilience to adaptive attack.

\paragraph{Defender} The defender sees the training images and the labels
attached to them. Nothing else: no clean subset, no poison mask, no trusted
held-out data, no knowledge of which classes are involved or what the trigger
looks like, and no signal that the dataset was attacked at all. The defender
may run a public pretrained encoder over the images, which reads pixels and
never reads a label~\cite{oquab2024dinov2}. It acts once, before training, by
deleting the examples it finds suspicious.

The defenses usually brought to this setting read the representations of a
model trained on the poisoned data
instead~\cite{tran2018spectral,chen2019activation,peri2020deepknn,hayase2021spectre}.
That model was fitted to the labels the defense is trying to question, and the
rest of this paper measures what follows from that.

\section{Audits: statistics and spaces}
\label{sec:audits}

A label audit has three parts a defender chooses independently. The three parts are numbered 3 to 5 in Figure~\ref{fig:overview} and are: the space the features live in, the statistic computed on top of them, and the rule that turns scores into deletions. All three are set in this section. Each detector below takes the same two inputs, a $N \times D$ matrix of $L_2$-normalised features and the $N$ assigned labels $\tilde{y}$, and nothing else. None of them ever see a true label, clean subset or poison mask.

\subsection{Two families of statistic}

\emph{Neighborhood} statistics ask whether a sample's label agrees with its
neighbors. It is the oldest rule in this family, going back to edited nearest
neighbors~\cite{wilson1972enn}. Writing $\mathcal{N}_k(i)$ for the $k$ nearest
neighbors of $i$ by cosine similarity, with $i$ itself excluded,
\begin{equation}
  d(i) \;=\; \frac{1}{k} \sum_{j \in \mathcal{N}_k(i)}
             \mathbf{1}\!\left[\tilde{y}_j \neq \tilde{y}_i\right].
  \label{eq:knn}
\end{equation}
We call this \texttt{knn\_disagreement}. It applies the criterion of Deep
$k$-NN~\cite{peri2020deepknn} as a continuous score rather than a vote, on
whichever space we hand it; the same neighborhood test is used to sanitise
labels under flipping attacks~\cite{paudice2019sanitization} and, in the
label-noise literature, to detect corrupted labels without training a
model~\cite{zhu2022simifeat}. \texttt{knn\_sim\_margin} is a continuous variant
that keeps the locality: mean similarity to the neighbors carrying the best
competing label, minus mean similarity to those carrying the assigned label.

\emph{Prototype} statistics~\cite{snell2017prototypicalnetworksfewshotlearning}
compare a sample to class centroids. Let $\mu_c$ be the normalised mean of the
features whose \emph{assigned} label is $c$. These centroids are built from the
labels the dataset carries, so the target-class centroid is contaminated by
whatever was flipped into it. This is the same contamination that affects the
class-conditional mean used by spectral signatures~\cite{tran2018spectral}, and
what motivates the robust estimators of SPECTRE~\cite{hayase2021spectre}. Using
clean labels here would be oracle access. Then
\begin{equation}
  p(i) \;=\; \max_{c} \langle f_i, \mu_c \rangle
             \;-\; \langle f_i, \mu_{\tilde{y}_i} \rangle,
  \label{eq:proto}
\end{equation}
which reads as ``how much better does some other class explain this image''.
\texttt{prototype\_trimmed} is our own robust answer to that contamination:
\eqref{eq:proto} with each centroid recomputed once after dropping the 20\% of
its class sitting furthest from a provisional centroid, since poisons are
exactly the members that sit far from the class they claim. The fraction is
fixed at 0.2 throughout and we never swept it (Section~\ref{sec:limitations}).

\paragraph{Score granularity} Equation~\eqref{eq:knn} is a count over $k$
neighbors, so at $k = 20$ it can only take 21 values. Over the 50\,000
CIFAR-10~\cite{krizhevsky2009cifar} training samples we measure exactly 21
distinct scores, and 396 of them
(truck$\rightarrow$deer; 156 on dog$\rightarrow$cat) sit at the maximum of 1.0.
Ties there are broken by a stable sort (that is, by dataset order), so the
detector does not really rank inside that group, and the group sits at the
suspicious end where the removal budget is spent.
Equation~\eqref{eq:proto} also ties heavily, but at the opposite end: 91.9\%
of samples on truck$\rightarrow$deer and 92.3\% on dog$\rightarrow$cat score
exactly zero, because their assigned class is already their best-matching
prototype. Those ties cost nothing, because no budget reaches that far down.
The part of the ranking a defender actually reads is spread over roughly 4\,000
distinct values. This is one reason the two statistics separate much further in
AUPRC than in AUROC, which we return to in Section~\ref{sec:claim2}.

\subsection{Four feature spaces}

We run the identical detectors on four spaces. Two are frozen and public:
DINOv2 ViT-S/14 (384-dimensional, 22.1\,M parameters) and ViT-B/14
(768-dimensional, 86.6\,M)~\cite{oquab2024dinov2}. Two are supervised: the
64-dimensional penultimate layer of a ResNet-32~\cite{he2016resnet} trained
on the poisoned labels
(the space Deep $k$-NN and SPECTRE actually
read~\cite{peri2020deepknn,hayase2021spectre}), and the same layer of the
same architecture trained on clean labels. Every image passes through its
encoder once under a deterministic evaluation transform with no augmentation,
and the features are $L_2$-normalised so that cosine similarity is an inner
product.

The DINOv2 spaces are label-independent in a precise sense: the encoder is
frozen, it is chosen before the data is seen, and no training label of this
dataset ever reaches it. The supervised-poisoned space is the opposite: it
is the product of fitting the labels under audit.

The clean-label space is not a defense, and we never present it as one: a
defender who already has clean labels has no need of an audit. It is in the
comparison as a control. It holds the architecture, the data, the statistic and
the metric fixed and changes only the labels the features were fitted to, which
separates ``supervised features are a poor space for this task'' from
``features fitted to the poison are a poor space for this task''. Section
\ref{sec:claim1} shows these two give very different answers.

\subsection{Turning scores into deletions}

Scores only matter once they become a removal set. We use three policies.
\texttt{auto:}$\tau$ deletes every sample scoring above $\tau$; it needs no
oracle, and at $\tau = 0.5$ for \eqref{eq:knn} it means ``most of your
neighbors disagree with your label''. This is the deployable operating point
and the one our headline mitigation numbers use. \texttt{topn:}$N$ deletes the
top $N$ by score, which lets us compare detectors at a matched budget so that a
difference in the trained model is about \emph{which} samples a detector chose
rather than how many. \texttt{perclass:}$Q$ deletes the top $Q$ fraction inside
each assigned class.

The per-class policy is not a convenience. Any detector whose score is
calibrated within a class is close to meaningless across classes, and a global
cut then spends the budget wherever that scale happens to run high. SPECTRE is
exactly such a detector, because it fits a separate robust covariance per
class. When we first gave it a global top-2\,728 cut on dog$\rightarrow$cat,
2\,372 of those removals landed outside the class holding 1\,499 of the 1\,500
poisons, and its recall read 0.233 where the same budget spent inside that
class reaches 0.805. That gap is a property of how we spent the budget, not of
SPECTRE. We therefore report SPECTRE under \texttt{perclass} throughout, which
is how it is meant to be deployed~\cite{hayase2021spectre}.

We also implement a cut at exactly the true number of poisons, because
precision@$k$ at that $k$ is what the Deep $k$-NN literature
reports~\cite{peri2020deepknn}. It requires the poison count, so we use it as a
diagnostic for comparability with prior work and never as an operating point.

\section{Experimental setup}
\label{sec:setup}

All experiments use CIFAR-10~\cite{krizhevsky2009cifar}: 50\,000 training
images and 10\,000 test images over ten classes. The victim is a
ResNet-32~\cite{he2016resnet} trained from scratch for 200 epochs with SGD,
learning rate 0.1, Nesterov momentum 0.9, weight decay $2 \times 10^{-4}$ and
batch size 256, with the learning rate cut by ten at epochs 75 and 125. Every
defended and undefended run in this paper uses that same recipe, so a
difference in the trained model comes from the data it was given and not from
how it was trained.

\begin{figure}[t]
  \centering
  \includegraphics[width=\columnwidth]{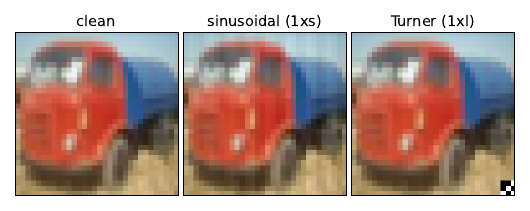}
  \caption{The two triggers, applied by the same code the experiments use. The
  sinusoidal trigger perturbs almost the whole image faintly (951 of 1\,024
  pixels, peak change 6/255); the Turner patch changes 9 pixels hard (peak
  199/255).}
  \label{fig:triggers}
\end{figure}

\paragraph{Attacks} We use the FLIP label sets at $m = 1500$, which is 3\% of
the training set. Two triggers appear in the
literature for this setting and we test both: a sinusoidal pattern added to the
whole image~\cite{barni2019sinusoidal} and a small high-contrast patch in the
image corner~\cite{turner2019labelconsistent}. Figure~\ref{fig:triggers} shows
both on a source-class image.

\paragraph{Class pairs} We attack four source$\rightarrow$target pairs:
truck$\rightarrow$deer, dog$\rightarrow$cat, car$\rightarrow$truck and
cat$\rightarrow$dog. The truck$\rightarrow$deer labels are the ones released
with FLIP; the other three we generated locally with the same pipeline.

Where these four sit among all CIFAR-10 pairs decides how the results should be
read, so we state it here. Section~\ref{sec:claim3} computes a stealthy-flip
pool for all 90 ordered pairs. Ranked by that pool, dog$\rightarrow$cat is
first (265), cat$\rightarrow$dog third (82) and car$\rightarrow$truck fourth
(72). The fifth-largest pool is 43, so our three non-zero pairs are the cluster
where the attacker has the most room and the audit should do worst.
truck$\rightarrow$deer is the other extreme, with a pool of exactly zero, and
it is the one pair we did not pick, since its labels come from FLIP.

Two of the four are the same class pair in both directions, which lets us ask
whether direction matters. It matters for the geometry and not for the victim:
the pool is 265 one way and 82 the other, while defended PTA is
$24.3 \pm 2.5$ and $24.4 \pm 2.7$ (Table~\ref{tab:predictor}).

\paragraph{Encoders} The audits use frozen DINOv2 ViT-S/14 throughout, with
ViT-B/14 as a scale ablation~\cite{oquab2024dinov2}. Features are cached once
per encoder and reused by every pair, attack and detector.

\paragraph{Seeds} Every load-bearing number is reported over three training
seeds. Detection metrics are deterministic given a feature cache and a label
set, so they carry no seed spread; only quantities read off a trained model do.

\begin{table}[t]
  \centering
  \caption{Reproducing FLIP on our setup. Published values are FLIP's Table~1,
  CIFAR-10 with a ResNet-32. Ours are the same configuration run here; the
  sinusoidal row is three seeds, the others are single runs. The clean and
  sinusoidal rows reproduce. The Turner row does not, and
  Section~\ref{sec:setup} says why.}
  \label{tab:repro}
  \begin{tabular}{llcc}
    \toprule
    setting & source & CTA & PTA \\
    \midrule
    \multirow{2}{*}{clean labels ($m = 0$)}
      & published & 92.38 & 0.1 \\
      & ours      & 92.57 & 0.1 \\
    \addlinespace
    \multirow{2}{*}{sinusoidal, $m = 1500$}
      & published & 89.87 & 99.8 \\
      & ours      & $89.47 \pm 0.35$ & $99.73 \pm 0.15$ \\
    \addlinespace
    \multirow{2}{*}{Turner, $m = 1500$}
      & published & 89.91 & 99.9 \\
      & ours      & 89.94 & \textbf{32.4} \\
    \bottomrule
  \end{tabular}
\end{table}

\paragraph{Reproduction} Table~\ref{tab:repro} compares our runs against the
numbers FLIP reports for the same configuration. The clean baseline and the
sinusoidal attack both land within a point of the published values, which is
what licenses everything that follows: the attack we defend against is the
attack the original paper describes.

The Turner row is a different story. Clean accuracy reproduces almost exactly
(89.94 against 89.91), so the pipeline is not at fault: the attack itself is
weak here, at PTA 32.4 against a published 99.9. We ran the single-corner
trigger of Figure~\ref{fig:triggers} against a four-corner published result, so
this is not a like-for-like comparison, and Section~\ref{sec:limitations} says
what we do with it. The sinusoidal trigger carries every claim in this paper.

\section{The feature space decides}
\label{sec:claim1}

The experiment in this section is a controlled swap. We fix the attack, the
labels, the detector and the metric, and change only the space the detector
reads. If the statistic were doing the work, the four spaces would give similar
answers. They do not.

\begin{table}[t]
  \centering
  \caption{AUPRC of every detector in every feature space, reported as
  truck$\rightarrow$deer / dog$\rightarrow$cat. Same labels, same detectors,
  same metric; only the space changes. Reading across a row isolates the
  effect of the space. The \emph{clean} column is an oracle control, not a deployable defense.}
  \label{tab:spaces}
  \setlength{\tabcolsep}{3pt}
  \small
  \begin{tabular}{lcccc}
    \toprule
    & \multicolumn{2}{c}{supervised} & \multicolumn{2}{c}{frozen DINOv2} \\
    \cmidrule(lr){2-3}\cmidrule(lr){4-5}
    detector & poisoned & clean & ViT-S/14 & ViT-B/14 \\
    \midrule
    \texttt{knn\_disagreement}  & .26/.05 & .97/.86 & .61/.35 & .74/.52 \\
    \texttt{knn\_sim\_margin}   & .25/.05 & .97/.88 & .63/.42 & .75/.59 \\
    \texttt{prototype\_margin}  & .22/.04 & 1.0/1.0 & .97/.67 & .99/.85 \\
    \texttt{prototype\_trimmed} & .27/.04 & 1.0/1.0 & .98/.73 & .99/.87 \\
    \bottomrule
  \end{tabular}
\end{table}

Table~\ref{tab:spaces} gives the result. Inside any one space the four
detectors sit close together; the same detector moves far more when the space
changes. The spread between the groups is larger than the spread within them,
which is the whole claim in one line. In the victim's own
representation every detector is close to useless. The
neighborhood audit reaches AUROC 0.546 on dog$\rightarrow$cat (chance is
0.5), and the prototype audit does no better at 0.511. Moving the same
detectors onto frozen DINOv2 features, with nothing else changed, takes
\texttt{knn\_disagreement} to AUROC 0.967 and \texttt{prototype\_trimmed} to
AUPRC 0.727. The ordering of the detectors is roughly preserved inside each
space, but the gap between spaces is far larger than the gap between
statistics.

\paragraph{The control that explains it} A sceptic can read the first column
two ways: supervised penultimate features may simply be a poor space for
finding mislabelled data, or they may be poor \emph{because} they were fitted
to the labels under audit. The clean-label column separates these. It is the
same architecture, the same images and the same 64 dimensions, trained with the
same recipe. Only the labels it fitted were clean. There the detectors are
near-perfect: AUPRC 0.966 and 1.000 on truck$\rightarrow$deer, 0.864 and 1.000
on dog$\rightarrow$cat. Supervised features are an excellent space for this
task. They stop being one once they are fitted to the poison. This is the
mechanism behind the whole paper, and it is why we frame the result as being
about label-independence rather than about self-supervision.

We should be clear about what that column is not. A defender who has clean
labels has no reason to audit anything, so this is a bound, not a method. Its
job is to rule out the alternative explanation, and it does.
\paragraph{The cheapest baseline: the victim's own loss} If the victim is
already trained, why pay for an encoder at all? Reading the model's loss on
each training sample costs nothing extra, and it is the standard way to find
bad labels~\cite{pleiss2020aum,han2018coteaching}. We tried it. Scoring by the
cross-entropy of the assigned label gives AUROC 0.893 and AUPRC 0.355 on
truck$\rightarrow$deer, and 0.757 / 0.102 on dog$\rightarrow$cat. Scoring by
the logit margin instead changes nothing (0.891 / 0.333 and 0.766 / 0.107).
The cost tells the same story: to catch 90\% of the poisons, the loss filter
deletes 14\,382 clean images on truck$\rightarrow$deer and 29\,393 on
dog$\rightarrow$cat. The prototype audit deletes 69 and 2\,082.

One detail is worth keeping. On dog$\rightarrow$cat the loss filter (AUPRC
0.102) still beats Deep $k$-NN on the same victim (0.049). The model's loss
remembers a little about which labels it was forced to fit. Its representation
remembers almost nothing, because the representation is what changed until
those labels fit. That is the mechanism of Table~\ref{tab:spaces} again, from
a second angle, with no new model trained.

\paragraph{Significance} The two spaces score the same 50\,000 samples, so the
comparison is paired. A paired bootstrap over samples ($B = 1000$) puts the
AUPRC gain of DINOv2 ViT-S/14 over the poisoned-model space, using
\texttt{prototype\_trimmed}, at $+0.710$ with a 95\% CI of $[+0.688, +0.730]$
on truck$\rightarrow$deer and $+0.686$, $[+0.661, +0.708]$, on
dog$\rightarrow$cat, both at $p < 10^{-3}$.

\paragraph{Encoder scale} A larger encoder improves every cell of
Table~\ref{tab:spaces} and reorders none of them: ViT-B/14 raises
\texttt{prototype\_trimmed} by $+0.018$ AUPRC on truck$\rightarrow$deer
($[+0.013, +0.022]$) and $+0.143$ on dog$\rightarrow$cat
($[+0.126, +0.159]$). The conclusion does not depend on which DINOv2 we use,
and the harder pair benefits more.

\paragraph{Does it produce a better model?} Detection is only worth reporting
if the trained model changes, so we retrain the victim on the filtered data.
Table~\ref{tab:victim} gives the result over three seeds. On
truck$\rightarrow$deer the attack is removed outright, and clean accuracy
recovers from 89.47 to 92.23, within half a point of the 92.57 a clean-label
model reaches, so the audit returns most of the accuracy the attack cost.
On dog$\rightarrow$cat the attack is cut from 94.67 to 24.30 but not removed.
We report that as a partial defense and return to why in
Section~\ref{sec:claim3}.

\subsection{What the alternative costs}
\label{sec:spectre}

\begin{table}[t]
  \centering
  \caption{Victim-level comparison, mean $\pm$ sd over three seeds. ``cut'' is
  how many of the 50\,000 training images the defense deletes. SPECTRE is given
  the per-class budget it is designed for (Section~\ref{sec:audits}); FLIP
  report it at CTA 91.47 / PTA 1.7 on truck$\rightarrow$deer~\cite{jha2023flip}.}
  \label{tab:victim}
  \setlength{\tabcolsep}{3pt}
  \small
  \begin{tabular}{lrcc}
    \toprule
    defense & cut & CTA & PTA \\
    \midrule
    \multicolumn{4}{l}{\emph{truck}$\rightarrow$\emph{deer}} \\
    none                        & 0       & $89.47 \pm 0.35$ & $99.73 \pm 0.15$ \\
    SPECTRE, $Q{=}0.20$         & 10\,000 & $90.94 \pm 0.23$ & $0.97 \pm 0.46$ \\
    \texttt{prototype\_trimmed} & 2\,981  & $\mathbf{92.23 \pm 0.14}$ & $\mathbf{0.20 \pm 0.20}$ \\
    \addlinespace
    \multicolumn{4}{l}{\emph{dog}$\rightarrow$\emph{cat}} \\
    none                        & 0       & $89.33 \pm 0.30$ & $94.67 \pm 0.42$ \\
    SPECTRE, $Q{=}0.20$         & 10\,000 & $89.12 \pm 0.29$ & $51.30 \pm 0.89$ \\
    \texttt{prototype\_trimmed} & 2\,728  & $\mathbf{91.88 \pm 0.18}$ & $\mathbf{24.30 \pm 2.51}$ \\
    \bottomrule
  \end{tabular}
\end{table}

The honest comparison for this paper is not against nothing. It is against
SPECTRE~\cite{hayase2021spectre}, the defense FLIP's own authors endorse after
testing three. So the first question is whether our reimplementation is strong
enough to be worth beating. It is: given the per-class budget it was designed
for, it reaches CTA $90.94 \pm 0.23$ and PTA $0.97 \pm 0.46$ on
truck$\rightarrow$deer, against the 91.47 / 1.7 that FLIP report for it in
their Table~16~\cite{jha2023flip} (Table~\ref{tab:victim}). We reproduce the published defense, and the rest of
this subsection prices it rather than disputing it.

The price is the deletion budget. SPECTRE reaches that result by removing 20\%
of every class, or 10\,000 images. The audit reaches a better result by
removing 2\,981, a 3.4$\times$ smaller cut, and keeps 1.3 more points of clean
accuracy. On dog$\rightarrow$cat the gap is wider on both axes: at the same
10\,000-image budget SPECTRE still leaves PTA 51.30, where the audit leaves
24.30 after deleting 2\,728.

\begin{figure*}[t]
  \centering
  \includegraphics[width=\textwidth]{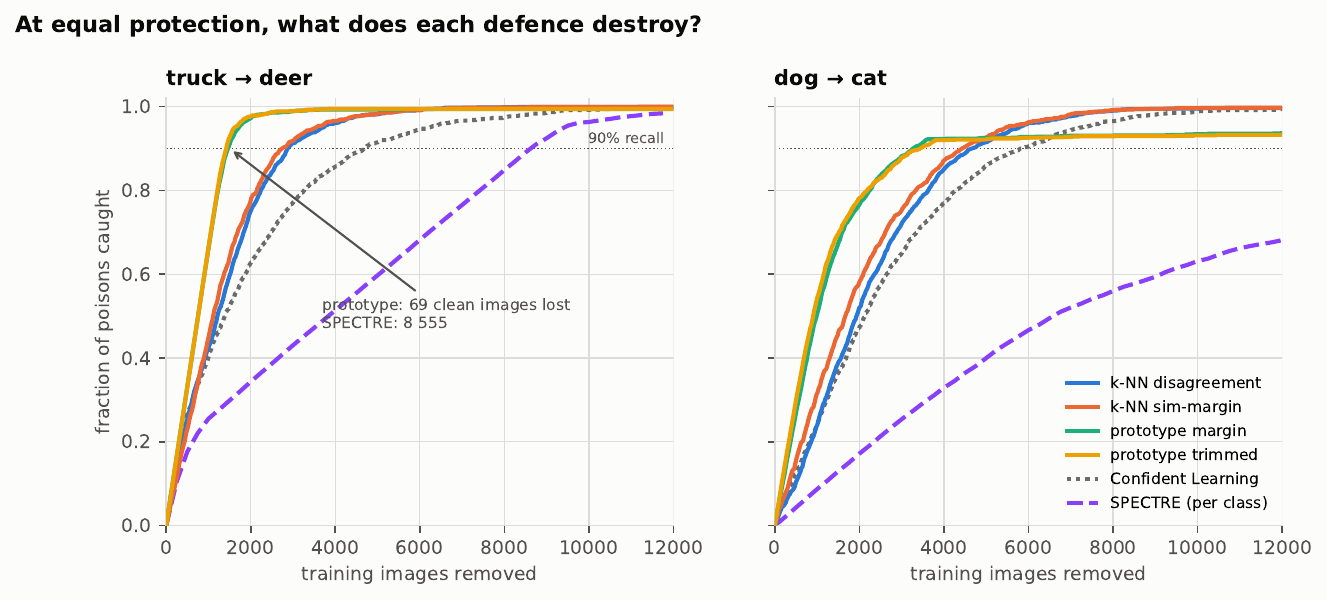}
  \caption{Protection against its price. At 90\% recall on
  truck$\rightarrow$deer a prototype audit on frozen features loses 69 clean
  images; SPECTRE, deployed per class as designed, loses 8\,555.}
  \label{fig:cost}
\end{figure*}

Figure~\ref{fig:cost} states this in the currency a defender actually budgets
in (clean images destroyed at equal protection), because recall at a fixed
budget flatters whoever chose the budget. To catch 90\% of the poisons on
truck$\rightarrow$deer, the prototype audit deletes 1\,419 images of which 69
are clean. SPECTRE needs $Q = 0.20$ to reach the same recall, and destroys
8\,555 clean images doing it: two orders of magnitude more, for the same
protection.

\begin{table}[t]
  \centering
  \caption{What each published defense deletes to catch 90\% of the poisons,
  deployed per class as all three are designed to be. Detection only, no
  retraining, so these sit alongside Figure~\ref{fig:cost}.}
  \label{tab:family}
  \setlength{\tabcolsep}{4pt}
  \small
  \begin{tabular}{lrr}
    \toprule
    defense & removed & clean destroyed \\
    \midrule
    \multicolumn{3}{l}{\emph{truck}$\rightarrow$\emph{deer}} \\
    SPECTRE, $Q{=}0.20$~\cite{hayase2021spectre}                & 10\,000 & 8\,555 \\
    activation clustering, $Q{=}0.20$~\cite{chen2019activation} & 10\,000 & 8\,630 \\
    spectral signatures, $Q{=}0.25$~\cite{tran2018spectral}     & 12\,500 & 11\,145 \\
    \texttt{prototype\_trimmed}                                 & 1\,419  & $\mathbf{69}$ \\
    \addlinespace
    \multicolumn{3}{l}{\emph{dog}$\rightarrow$\emph{cat}} \\
    \multicolumn{3}{l}{\quad SPECTRE: never reaches 90\% (0.849 at $Q{=}0.50$)} \\
    \multicolumn{3}{l}{\quad spectral signatures: never reaches 90\% (0.764 at $Q{=}0.50$)} \\
    activation clustering, $Q{=}0.50$~\cite{chen2019activation} & 25\,001 & 23\,617 \\
    \texttt{prototype\_trimmed}                                 & 3\,432  & $\mathbf{2\,082}$ \\
    \bottomrule
  \end{tabular}
\end{table}

\paragraph{The cost is the family's, not SPECTRE's} SPECTRE is the defense FLIP
endorses, but it is one of three in the same line, so we ran the other two on
the same features under the same budget rule. On truck$\rightarrow$deer all
three land in one band: 8\,555 clean images for SPECTRE, 8\,630 for activation
clustering, 11\,145 for spectral signatures. The audit reaches the same recall
for 69. On dog$\rightarrow$cat two of the three never reach 90\% recall at all,
even after deleting half of every class, and the third needs $Q = 0.50$ and
23\,617 clean images where the audit needs 2\,082. The deletion cost priced in
this section belongs to searching the poisoned model's representation, not to
one method.

One caveat belongs here rather than in a footnote. That comparison is the
realistic deployment, in which SPECTRE has to find the poisoned class itself.
If it is instead handed the target class (an oracle no defender has), it
becomes much more efficient, and on dog$\rightarrow$cat it is roughly
competitive with us: 0.828 recall for 1\,739 clean images destroyed, against
our 0.900 recall for 2\,082. The one-sided result is on
truck$\rightarrow$deer; on the harder pair the honest statement is narrower.

\paragraph{Where our claim stops} These two defenses also mark the edge of
Section~\ref{sec:claim1}. Moving them to the frozen encoder does not help, and
on dog$\rightarrow$cat it hurts: inside the target class, activation clustering
scores AUROC 0.903 on the poisoned model's features and 0.122 on DINOv2.
Neither method reads a label. Both look for a minority cluster or a stretched
direction inside an assigned class, and the poisoned model is what builds that
structure, because it has learned the trigger. A frozen encoder has no trigger
to respond to, so on a pair whose classes already overlap the 2-means split
inside the target class comes out 47.7/52.3 and says nothing about the poison.
Our claim covers statistics that compare a sample against its assigned label.
It does not cover statistics that ignore labels, and these runs are what draw
that line.

\section{Static detection quality selects the wrong audit}
\label{sec:claim2}

Section~\ref{sec:claim1} ranked audits by how well they retrieve poisons.
That is how the literature reports them, and this section is about why it is
not enough. We give three ways the static ranking misleads: it does not
predict the trained model, it does not survive a change of metric, and it
inverts once the attacker knows which audit is running.

\begin{table}[t]
  \centering
  \caption{Every detector at the same removal budget (2\,981 and 2\,728
  images), so a difference in the model is about \emph{which} samples were
  cut. Detection and outcome disagree repeatedly: the arrows mark pairs of
  rows where the better-ranking detector produces the worse model.}
  \label{tab:detectors}
  \setlength{\tabcolsep}{3pt}
  \small
  \begin{tabular}{lcccc}
    \toprule
    detector & AUROC & AUPRC & CTA & PTA \\
    \midrule
    \multicolumn{5}{l}{\emph{truck}$\rightarrow$\emph{deer}} \\
    \texttt{knn\_disagreement}  & .986 & .614 & 91.99 & 8.6 \\
    \texttt{knn\_sim\_margin}   & .987 & .630 & 92.04 & 11.4 $\downarrow$ \\
    \texttt{prototype\_margin}  & .996 & .968 & 92.34 & 0.4 \\
    \texttt{prototype\_trimmed} & .996 & .976 & 92.31 & \textbf{0.2} \\
    cleanlab~\cite{northcutt2021confident} & .974 & .584 & 91.48 & 51.3 \\
    \addlinespace
    \multicolumn{5}{l}{\emph{dog}$\rightarrow$\emph{cat}} \\
    \texttt{knn\_disagreement}  & .967 & .348 & 90.51 & 53.6 \\
    \texttt{knn\_sim\_margin}   & .971 & .424 & 90.72 & 56.9 $\downarrow$ \\
    \texttt{prototype\_margin}  & .951 & .673 & 91.74 & \textbf{23.0} \\
    \texttt{prototype\_trimmed} & .951 & .727 & 92.04 & 25.7 $\downarrow$ \\
    cleanlab~\cite{northcutt2021confident} & .959 & .353 & 90.57 & 67.3 \\
    \bottomrule
  \end{tabular}
\end{table}

\subsection{Better detection does not mean a better model}

Table~\ref{tab:detectors} gives every detector at a matched removal budget, so
the only thing that varies is which samples each one chose. If detection
quality were a proxy for protection, AUPRC and PTA would move together. They
do not.

\texttt{knn\_sim\_margin} is the continuous variant of
\texttt{knn\_disagreement} and beats it on both metrics and both pairs. At the
victim it loses on both pairs, 11.4 against 8.6 and 56.9 against 53.6. The
same reversal appears inside the prototype family: on dog$\rightarrow$cat
\texttt{prototype\_trimmed} retrieves more poisons (AUPRC 0.727 against 0.673)
and leaves a slightly worse model (25.7 against 23.0). We had originally read
the first of these as an improvement, on the strength of its detection score
alone. It is not one.

\paragraph{The same point with a published method} A reader might grant this
for two of our own statistics and still trust the ranking across methods, so we
ran Confident Learning~\cite{northcutt2021confident} on the same frozen
features, giving it cross-validated probabilities, a more generous setting
than it usually gets. On dog$\rightarrow$cat it attains a \emph{higher} AUROC
than our audit (0.959 against 0.951) while retrieving roughly half the poisons
(AUPRC 0.353 against 0.727), and the model it produces leaves PTA 67.3 against
25.7. The pattern holds on car$\rightarrow$truck as well (AUROC 0.961 against
0.959, AUPRC 0.480 against 0.789); on truck$\rightarrow$deer and
cat$\rightarrow$dog the AUROC ordering is correct, though on cat$\rightarrow$dog
it is a 0.001 gap (0.969 against 0.970) beside an AUPRC gap of 0.29 (0.494
against 0.786). So on two of four pairs, a
reader ranking these detectors by the metric the literature reports would pick
the one that leaves more than twice the attack success. AUROC is not merely
pessimistic here; it is unreliable in direction.
Across all four pairs the same thing happens at the detection level: the four
AUROC gaps are nearly level and cross twice, while the same four AUPRC gaps are
wide and never cross. The metric the literature reports is the one that cannot
tell these detectors apart.

The mechanism is the one described in Section~\ref{sec:audits}: AUROC is
dominated by the easy bulk of the ranking, and a removal budget only ever
touches its head.

\begin{figure}[t]
  \centering
  \includegraphics[width=\columnwidth]{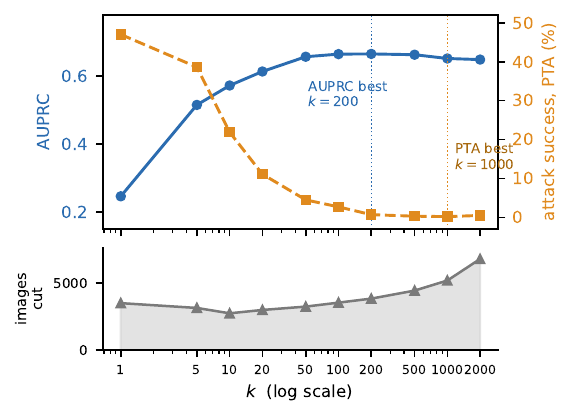}
  \caption{Twenty victim trainings sweeping $k$ on truck$\rightarrow$deer.
  Detection peaks at $k = 200$ while the trained model keeps improving past it.
  The lower panel is the reason the curves separate: at the \texttt{auto}
  threshold $k$ also sets how much data is cut.}
  \label{fig:ksweep}
\end{figure}

\paragraph{Tuning a hyperparameter on the same score} The disagreement also
shows up when choosing $k$. Figure~\ref{fig:ksweep} sweeps it over twenty
victim trainings. AUPRC peaks at $k = 200$, but the model keeps getting better
after that: $k = 500$ reaches both a higher clean accuracy (92.36 against
92.27) and a lower attack success (0.3 against 0.7). A defender who tuned $k$
on the detection score would stop one step early. We should be careful about
the mechanism here, because it is not the same as above. At the \texttt{auto}
threshold $k$ controls both the statistic and the size of the cut, so a larger
$k$ deletes more: 3\,827 images at $k = 200$ against 4\,425 at $k = 500$.
AUPRC charges for that extra deletion and protection does not. This is a budget
effect rather than a ranking effect, and it is a weaker instance of the claim
than the two above, but it points the same way.

\subsection{The statically better audit breaks first}

\begin{figure}[t]
  \centering
  \includegraphics[width=\columnwidth]{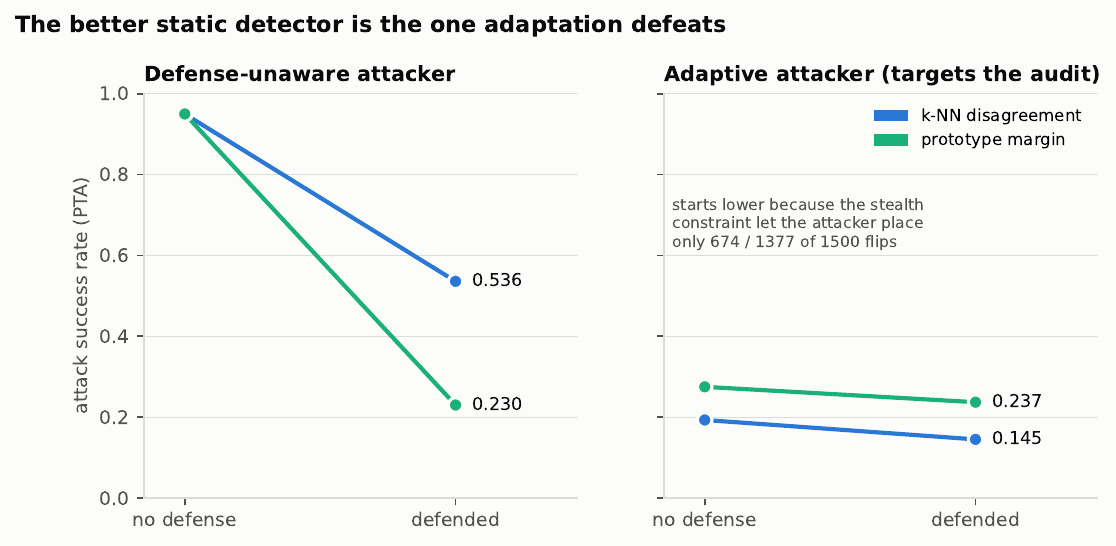}
  \caption{The statistic that wins under a defense-unaware attacker is the one
  a defense-aware attacker dismantles first.}
  \label{fig:reversal}
\end{figure}

The ranking also inverts once the attacker knows the audit. We let the attacker
choose which labels to flip subject to staying below the defender's score, with
the defender holding a fixed false-positive budget, and we give attacker and
defender the same scoring code so the two cannot drift apart
(Section~\ref{sec:background}).

Against an attacker that targets it, the prototype audit, the winner of every
static comparison above, retrieves 1.4\% of the placed flips on
dog$\rightarrow$cat. The coarse neighborhood audit, which loses every static
comparison, retrieves 58.5\% of the same flips. Figure~\ref{fig:reversal}
shows the reversal as a swap: the two statistics change places between the left
panel, where the attacker ignores the audit, and the right panel, where it
targets one.

The coarse statistic has a second advantage that recall alone hides: evading it
is expensive. At the same stealth budget the attacker can place only 5 flips
that stay below the neighborhood score, against 579 below the prototype score.
A saturating count is hard to slip past precisely because it does not offer the
attacker a fine gradient to descend.

\subsection{Auditing with an untargeted statistic does not rescue it}

\begin{figure}[t]
  \centering
  \includegraphics[width=\columnwidth]{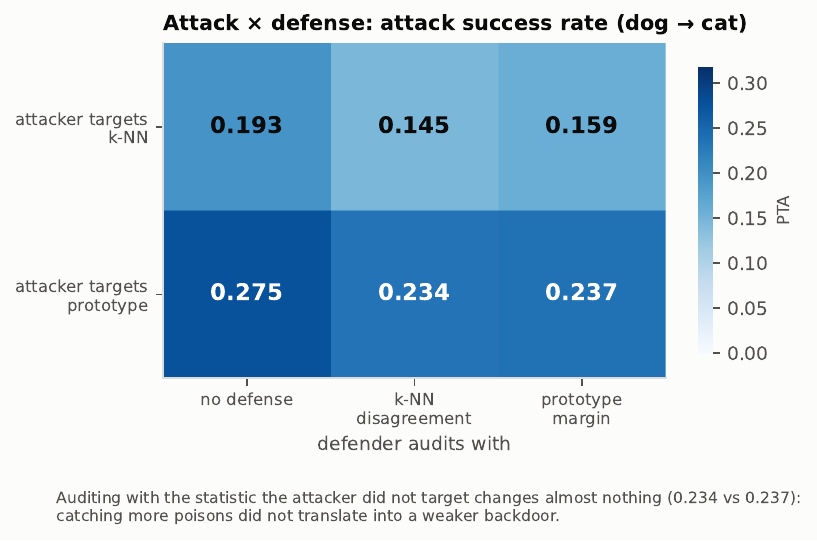}
  \caption{Attack $\times$ defense, dog$\rightarrow$cat, three seeds.}
  \label{fig:matrix}
\end{figure}

If one audit is evaded, the obvious move is to audit with a different one. We
tested it directly: the attacker targets one statistic, the defender audits
with either the same statistic or the other, and both cut 5\% of the data.
Figure~\ref{fig:matrix} and the following numbers are over three seeds. What
the figure shows is an absence: the off-diagonal bars do not sit below the
diagonal ones and the error bars overlap, so switching statistics buys nothing
visible.

The answer is no. Against the attacker that targets the neighborhood audit,
switching to the prototype audit makes the model \emph{worse}
($15.77 \pm 0.15$ against $13.77 \pm 0.81$ PTA, $p = 0.046$). Against the
attacker that targets the prototype audit, switching to the neighborhood audit
gives $22.20 \pm 1.74$ against $24.17 \pm 0.40$, which three seeds cannot
resolve ($p = 0.185$). Removal itself still helps, buying 4 to 5 PTA points
over no defense in both columns, but which statistic does the removing
stops mattering once the attacker has adapted.

\paragraph{Removal saturates} The same experiment shows why. In the
prototype-targeted column the off-diagonal defender catches 45.0\% of the
placed flips and the matched defender catches 28.4\%, a 1.6$\times$ difference
in recall. The models they produce differ by about two PTA points, and not
resolvably. Past some point, catching more poisons stops buying protection,
which is the second reason a detection score is a poor summary of a defense.

\section{Predicting where the audit applies}
\label{sec:claim3}

Sections~\ref{sec:claim1} and~\ref{sec:claim2} show that the audit removes the
attack on one pair and only halves it on another. A defender would like to know
which case they are in \emph{before} committing to the defense. This section
asks whether the encoder alone answers that, and the answer is yes, but only
as a yes-or-no, not as a degree.

\paragraph{The stealthy-flip pool} For an ordered pair
$a \rightarrow b$, take every training image of class $a$ and ask what
score the audit of \eqref{eq:knn} would give it \emph{if} it were relabelled to
$b$. That score is one minus the fraction of its neighbors already carrying
label $b$. The \emph{pool} at threshold $\tau$ is the number of class-$a$
images whose hypothetical score would fall at or below $\tau$ (that is, how
many flips the attacker could place without tripping the audit). We report it at
$\tau = 0.5$.

The pool needs the encoder and the clean labels, and nothing else. No attack,
no poisoned labels, no training run. It takes seconds once the features are
cached, and we computed it for all 90 ordered CIFAR-10 pairs before running any
of the attacks in this paper.

\begin{table}[t]
  \centering
  \caption{The pool against what happened. Pool sizes come from encoder
  geometry alone; AUPRC is \texttt{prototype\_trimmed}; PTA columns are mean
  $\pm$ sd over three seeds at each pair's matched budget. Detection tracks the
  pool; the trained outcome only separates zero from non-zero.}
  \label{tab:predictor}
  \setlength{\tabcolsep}{4pt}
  \small
  \begin{tabular}{lrccc}
    \toprule
    pair & pool & AUPRC & PTA none & PTA defended \\
    \midrule
    truck$\rightarrow$deer & 0   & .976 & $99.7 \pm 0.2$ & $\mathbf{0.2 \pm 0.2}$ \\
    car$\rightarrow$truck  & 72  & .789 & $99.4 \pm 0.3$ & $24.6 \pm 1.3$ \\
    cat$\rightarrow$dog    & 82  & .786 & $94.4 \pm 0.7$ & $24.4 \pm 2.7$ \\
    dog$\rightarrow$cat    & 265 & .727 & $94.7 \pm 0.4$ & $24.3 \pm 2.5$ \\
    \bottomrule
  \end{tabular}
\end{table}

\paragraph{What it predicts} Table~\ref{tab:predictor} gives the four pairs we
attacked. On
truck$\rightarrow$deer the pool is exactly zero: no source-class image can be
relabelled to deer and still look ordinary to its neighbors. The attack there
is still near-perfect undefended, at PTA 99.7, but the audit removes it
completely, leaving $0.2 \pm 0.2$. On the other three pairs a stealthy pool exists and the audit leaves roughly a
quarter of the attack standing. The spreads in Table~\ref{tab:predictor} carry
the point its ordering cannot: the three non-zero pools land in one band and
overlap, so this is a step, not a slope.

\begin{figure}[t]
  \centering
  \includegraphics[width=\columnwidth]{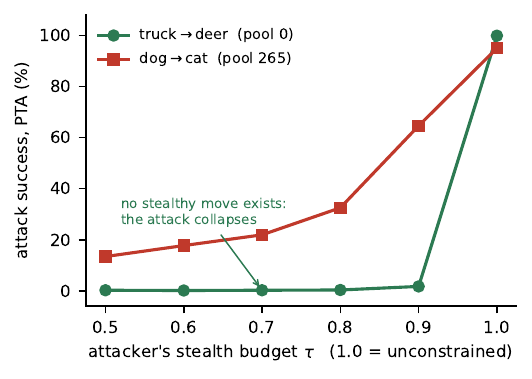}
  \caption{What the pool means for the attacker. As the stealth budget $\tau$
  tightens, dog$\rightarrow$cat degrades smoothly while
  truck$\rightarrow$deer collapses as soon as any constraint applies: with a
  pool of zero there is no flip that both installs the backdoor and survives
  the audit.}
  \label{fig:frontier}
\end{figure}

Figure~\ref{fig:frontier} shows what the pool means from the attacker's side.
Tighten its score budget at all and the zero-pool pair has nowhere to go: attack
success falls from 99.9 unconstrained to 1.7 at $\tau = 0.9$ and 0.2 at
$\tau = 0.5$. The other pair trades stealth for strength smoothly instead. The
collapse on truck$\rightarrow$deer is therefore structural, because the attacker has
no move, rather than evidence that the defense is strong there.

That separation is sharp and it was predicted in advance, but it rests on one
pair. Twenty-nine of the ninety ordered CIFAR-10 pairs have a zero pool, and we
trained victims for one of them, so the zero side of the binary is a single
measurement. What the pool also does \emph{not} do is order the three pairs
where an attack exists. Pools of 72, 82
and 265 (a 3.7$\times$ range) give defended PTA of $24.6 \pm 1.3$,
$24.4 \pm 2.7$ and $24.3 \pm 2.5$. Those three numbers sit within 0.3 points of
each other, with spreads that overlap entirely. There is no ordering here to
report, and three seeds are enough to say so rather than to leave it open.

\paragraph{Why the gap is the paper's own result} The pool does order something.
It orders \emph{detection}, monotonically and on all four pairs: AUPRC 0.976,
0.789, 0.786, 0.727 as the pool grows. So the predictor behaves exactly as
designed at the detection level and then flattens at the model level, which
is the negative result of Section~\ref{sec:claim2} reappearing inside our own
predictor. Better detection does not imply a better model, and a quantity built
to forecast detection therefore forecasts the model only coarsely. We state
claim 3 as a binary for that reason, and we do not fit a curve through four
points.

\paragraph{A caveat on the measurement} The pool is computed against clean
neighbor labels, so it is a lower bound. Once many flips land in the same
region of feature space they begin to shield one another, and the attacker's
real room is larger than the count suggests. This does not affect the binary
reading, since a pool of zero stays zero, but it does mean the numbers should
not be read as an attacker's exact budget.

\section{Cost}
\label{sec:cost}

The obvious objection to auditing against a frozen encoder is that it replaces
a cheap defense with a large pretrained model. That has to be stated in numbers
rather than waved away, so we measured every stage on the real 50\,000-image
training set rather than estimating from FLOP counts.

\begin{table}[t]
  \centering
  \caption{Measured cost of each stage, against one victim training. Encoder
  timings are a single GPU run on a Tesla P6; the CPU rows and the victim
  median come from a separate run on the same cluster. The encoder is the only
  substantial new cost, and it is paid once per dataset.}
  \label{tab:cost}
  \setlength{\tabcolsep}{3pt}
  \small
  \begin{tabular}{lrr}
    \toprule
    stage & seconds & vs.\ training \\
    \midrule
    \multicolumn{3}{l}{\emph{paid once per dataset}} \\
    DINOv2 ViT-S/14 pass (22.1\,M) & 277.6 & 8.2\% \\
    DINOv2 ViT-B/14 pass (86.6\,M) & 1\,190.0 & 35.3\% \\
    $k$-NN graph, $k = 20$ & 36.4 & 1.1\% \\
    \addlinespace
    \multicolumn{3}{l}{\emph{paid per audit}} \\
    \texttt{prototype\_trimmed} & 0.09 & $<0.01$\% \\
    \texttt{knn\_disagreement} (graph cached) & 0.01 & $<0.01$\% \\
    SPECTRE & 2.2 & 0.07\% \\
    \addlinespace
    victim training, 200 epochs (median, $n{=}99$) & 3\,372 & 100\% \\
    \bottomrule
  \end{tabular}
\end{table}

Table~\ref{tab:cost} gives the result. The encoder pass dominates everything
downstream, and it is paid \emph{once per dataset}, not once per attack: the
features never touch a label, so a single cache served every class pair, every
attack and every detector in this paper. The audit we recommend does not even
need the $k$-NN graph, since prototypes are computed directly from the features.
Scoring 50\,000 samples then takes under a tenth of a second on a CPU.

\paragraph{The comparison that matters} Encoder-versus-nothing is the wrong
framing, because a defense has to be compared against what a defender would
otherwise do. SPECTRE and Deep $k$-NN read the representations of a model
trained on the poisoned data, so before either can score anything, the defender
must train that model, which is a full training run. Counting the whole pipeline, and
assuming the defender retrains on the filtered data in both cases:
\begin{itemize}
  \item frozen-encoder audit: encoder pass, scoring, retrain
        $= 3\,650$\,s, or $1.08$ training runs;
  \item SPECTRE: train victim, score, retrain
        $= 6\,746$\,s, or $2.00$ training runs.
\end{itemize}
So the audit costs roughly 8\% more than simply training the model once, while
the representation-based route costs twice that. A defender who suspects
poisoning and responds by retraining alone pays a full run and still has no
evidence about which samples were bad; the audit costs a fraction of one and
returns a ranked list.

\paragraph{What these numbers are worth} They come from one GPU on one shared
cluster, so they should be read as ratios rather than absolutes. The CPU
timings in particular move with machine load: an earlier measurement of the
same operations on the same cluster gave 31.2\,s for the graph and 1.56\,s for
SPECTRE, against the 36.4 and 2.2 reported here. The encoder row is a single
measurement, not a median. None of this changes the ordering, which is what the
argument rests on.

\section{Related work}
\label{sec:related}

\paragraph{Backdoor attacks} Backdoors began as a supply-chain threat in which
the attacker edits images and labels together~\cite{gu2017badnets,chen2017targeted}.
Most work since has gone into hiding the image edit: periodic
patterns~\cite{barni2019sinusoidal}, reflections~\cite{liu2020reflection},
warping~\cite{nguyen2021wanet}, learned triggers~\cite{doan2021lira},
feature-space triggers~\cite{saha2020hidden}, and grafted trojan
modules~\cite{tang2020trojannet}. Clean-label attacks keep the label correct and
perturb the image so the trigger becomes the easiest feature to
learn~\cite{turner2019labelconsistent,gao2023notall}, and
some triggers occur naturally~\cite{wenger2022natural}. Backdoors can also enter
through the loss~\cite{bagdasaryan2021blind} or be made cryptographically
undetectable~\cite{goldwasser2022undetectable}. Surveys cover the
area~\cite{li2022survey,cina2023wildpatterns}, and poisoning web-scale corpora
is now practical~\cite{carlini2024webscale}, which makes untrusted training data
a present concern~\cite{barreno2010security}.

\paragraph{Label-only attacks} Our threat model is FLIP's~\cite{jha2023flip}:
flipping labels alone suffices, using trajectory
matching~\cite{cazenavette2022trajectory} from dataset
distillation~\cite{wang2018datasetdistill,nguyen2021infinite,sucholutsky2021softlabel}.
A parallel result attacks multi-label models the same
way~\cite{chen2023cleanimage}. Outside deep vision, label flipping is studied as
poisoning of deployed security classifiers (malware~\cite{taheri2020defending,aryal2023labelflip},
URLs~\cite{nowroozi2024mitigating}, activity
recognition~\cite{shahid2022labelflip}) and defended by sanitising labels
against their neighborhoods~\cite{paudice2019sanitization} or filtering
federated clients~\cite{rahman2025antiflipper}. Recent work adds adversarial
training~\cite{bal2025adversarial} and exact certificates for some
architectures~\cite{sabanayagam2025certification}. Those certify or harden
training; we audit the labels before it.

\paragraph{Defenses that read the poisoned model} The closest line searches the
representations of a model trained on the poison. Spectral signatures use the
top singular direction~\cite{tran2018spectral}, activation clustering splits
each class in two~\cite{chen2019activation}, Deep $k$-NN checks neighborhood
label agreement~\cite{peri2020deepknn}, and SPECTRE adds robust estimation and a
quantum-entropy score~\cite{hayase2021spectre}. Newer members read activation
gradients~\cite{yuan2024activationgrad} or stabilise detection with
sharpness-aware minimisation~\cite{zhang2024sharpness}. All assume the poison stays separable in
a representation fitted to the poisoned labels, which an adaptive adversary can
break~\cite{qi2023latent,shokri2020bypassing}. We run the first two in
Section~\ref{sec:spectre} rather than only citing them. We keep the statistic
and change the space.

\paragraph{Defenses elsewhere in the pipeline} Others act on the model or at
inference: fine-pruning removes dormant units~\cite{liu2018finepruning}, Neural
Cleanse recovers the trigger by optimisation~\cite{wang2019neuralcleanse},
universal litmus patterns classify whole models~\cite{kolouri2020ulp}, and
STRIP, SentiNet and Scale-Up test inputs at run
time~\cite{gao2019strip,chou2020sentinet,guo2023scaleup}. Explanation-based
forensics compares a suspect model against a trusted clean one, and recovers
even sparse, imperceptible triggers from the difference between their Grad-CAM
attention maps~\cite{dey2026gradcam}, which needs both a trained model to
interrogate and a trigger in the image to react to. A label-only attack
supplies neither. Training-time methods decouple representation from
classifier~\cite{huang2022decoupling} or distil the trigger
away~\cite{li2021nad}. A separate line gives guarantees rather than
heuristics~\cite{steinhardt2017certified,weber2020rab,wang2020randomized},
though stronger attacks break sanitisation without
them~\cite{koh2022stronger}, and benchmarks show reported numbers move with
setup~\cite{schwarzschild2021toxic}. Poisoning need not use labels at all:
gradient matching~\cite{geiping2021witches} and corrupted
augmentation~\cite{rance2022augmentation}. A recent systematisation argues the
evaluation itself is the weak link, naming missing adaptive attackers and narrow
metrics~\cite{abad2025sok}; Section~\ref{sec:claim2} is an instance of that
concern, measured on one attack.

\paragraph{Label error detection and audit spaces} Our audit is, mechanically, a
label-error finder. Confident Learning is the standard method and our closest
published comparison~\cite{northcutt2021confident}; on identical features it
retrieves fewer poisons than its AUROC suggests (Section~\ref{sec:claim2}).
Loss- and margin-based filters do the same from the model's own
fit~\cite{pleiss2020aum,han2018coteaching}, and we evaluate one in
Section~\ref{sec:claim1}. The crowdsourcing literature treats noisy labels as
estimation~\cite{sheng2008getanother,karger2011crowdsourcing}, the setting
FLIP's first scenario assumes. Auditing against a frozen public encoder is
becoming common: we use DINOv2~\cite{oquab2024dinov2}, a vision
transformer~\cite{dosovitskiy2021vit} trained without labels, and concurrent
work finds backdoors in CLIP data by similar
reasoning~\cite{huang2025clipbackdoor}. What we add is not the encoder but the
controlled comparison: hold the audit fixed, vary only the space, with a
clean-label model as the control.

\paragraph{Knowledge distillation} FLIP's second scenario is
distillation~\cite{hinton2015distillation}, where a teacher's soft labels are
the only channel. Distillation has been proposed as a defense in its own
right~\cite{papernot2016distillation}, but backdoors survive
it~\cite{ge2021antidistillation}. We do not evaluate that scenario.

\section{What These Results Do Not Establish}
\label{sec:limitations}

Section~\ref{sec:setup} says how the experiments were run; this says what they
do not show. We have tried to leave nothing out.

\paragraph{Scope} One dataset, one architecture, one working trigger, one
budget: CIFAR-10, ResNet-32, the sinusoidal trigger, $m = 1500$. FLIP covers
three datasets and three architectures~\cite{jha2023flip}; our code supports all
of them and we ran one of each, so the gap is compute, not engineering. We have
no evidence about more classes, larger images, other architectures, or budgets
far from 3\%. The truck$\rightarrow$deer labels are the FLIP release; the other
three pairs we regenerated. The Turner trigger did not reproduce: we ran the
single-corner variant against a four-corner published result
(Section~\ref{sec:setup}), so we treat it as partial and draw no mitigation
conclusion. By stealthy-flip pool the four hardest of the 90 ordered pairs are
dog$\rightarrow$cat (265), truck$\rightarrow$car (83), cat$\rightarrow$dog (82)
and car$\rightarrow$truck (72); we ran three. The missing one,
truck$\rightarrow$car, is the inverse of car$\rightarrow$truck, and the one
untested direction among the hardest pairs. Each pair costs three victim
trainings, which is why we report this rather than fix it.

\paragraph{The encoder has already seen this kind of image} The limitation we
would raise first. DINOv2 reaches 96.3\% zero-shot $k$-NN accuracy on CIFAR-10,
so the audit is label-independent but not knowledge-independent: it never reads
a training label, yet the encoder was pretrained on web images that look like
CIFAR-10. On a domain it has not seen (medical scans, satellite imagery,
industrial inspection), the neighborhoods carrying our whole signal may not
form. We do not test that.

\paragraph{Measurement} \texttt{cudnn.benchmark} is on, so two runs of the same
configuration differ; the noise floor is $\approx$0.01 CTA and every
load-bearing cell is three seeds. Detection metrics are deterministic given a
feature cache and a label set, so they carry no spread, though saved scores
depend on TF32 matmul, which moves about 0.05\% of samples. Precision@$k$ sizes
its cut with the true poison count and is reported only for comparability with
prior work. Timings come from one GPU on one cluster: read them as ratios, not
absolutes.
\paragraph{Two choices a defender could not make} The trimming fraction of
\texttt{prototype\_trimmed} is 0.2 everywhere here, the first value we tried;
sweeping it costs seconds and we should have. Choosing $k$ is harder. On
truck$\rightarrow$deer, picking it by stability costs $0.0006$ AUPRC at the 3\%
budget and $0.0083$ at 5\% and 10\%; on dog$\rightarrow$cat no rule wins at
every budget, with agreement giving up $0.007$ and $0.011$ at the two smaller
budgets where stability gives up $0.106$, and the order reversing at 10\%,
where stability lands on the oracle and agreement gives up $0.069$ (Otsu picks
$k = 1$ everywhere and should not be used; Appendix~\ref{app:kselect}). So $k$-NN numbers on the harder pair carry
a tuning advantage a real defender would not have; the prototype audit has no
$k$. Same flavour in the cost comparison: the 90\%-recall cut in
Section~\ref{sec:spectre} uses the ground-truth mask, which is the right way to
compare detectors at equal protection but not an operating point. On
truck$\rightarrow$deer the deployable threshold catches 91.3\%, so the two
nearly coincide; on dog$\rightarrow$cat it catches 67.7\%, and 90\% is a point
the deployed audit never reaches.

\paragraph{Where the claims stop} Handed the target class (an oracle no
defender has), SPECTRE becomes far more efficient and is roughly competitive
with us on dog$\rightarrow$cat; our advantage holds in the realistic deployment
and on the easier pair, and we claim no more. Removal is the only mitigation we
test: we delete and retrain, never relabelling, down-weighting, a robust loss,
or the soft labels the attack itself produces. Since removal
saturates (Section~\ref{sec:claim2}), that is the obvious next thing to try. Our
adaptive attacker picks which labels to flip but not what to flip them to, and
does not re-optimise the experts under the stealth constraint, so every
adaptive number bounds the defense from above. The stealthy pool only separates
pairs where no stealthy attack exists from pairs where one does: it does not
order the second group, four pairs is a thin basis, and of the 29 zero-pool
pairs among the 90 we trained victims for exactly one, so that half of the
binary is a single observation. Computed against clean neighbor labels, it is
also a lower bound, since flips landing in the same region shield each other.

\section{Conclusion and future work}
\label{sec:conclusion}

We asked how much of a label audit's power comes from the statistic it computes
and how much from the space it computes it in. On this attack the space
decides. The same neighborhood audit that sits near chance on the poisoned
victim's own features (AUROC 0.546 on dog$\rightarrow$cat) reaches 0.967 on
frozen DINOv2 features, and the same architecture trained on clean labels
reaches 0.995, so the deficit belongs to fitting the poison, not to
supervised learning.

That reframing has a practical consequence. SPECTRE, the defense FLIP's authors
endorse, reproduces on our setup, and once reproduced it can be priced: it
needs to delete 20\% of every class to remove the backdoor, where an audit on
frozen features does better on both axes after deleting 6\%. The gain is not a
cleverer statistic. It is reading a space the attacker could not bend.

Two results cut against the easy version of this story, and we think they are
the more useful half. First, detection quality does not rank audits correctly:
the statistic that wins every static comparison is the one a defense-aware
attacker dismantles first, and on two of four class pairs a published label-error
method attains a higher AUROC than our audit while leaving more than twice the
attack success in the trained model. Second, removal saturates: past a point,
catching more poisons stops buying protection. Any future work in this area
should report the trained model, not only the ranking.

\paragraph{What we would do next} The cheapest gap to close is the trimming
fraction, which sits at its default in every run here and costs seconds to
sweep. The most obvious one is breadth: CIFAR-100 and Tiny-ImageNet with
ResNet-18 and VGG-19 are the settings the attack paper already covers, they are
supported by the code we used, and running them would say whether ``the space
decides'' survives more classes and larger images. That is a question about
generality. A different and more important question is about the claim itself:
DINOv2 already knows CIFAR-10 well, so a dataset outside its pretraining
distribution is the only way to tell label-independence apart from encoder
familiarity. Both are needed, and they are not the same experiment. Because removal saturates, the more promising direction for
mitigation is to stop deleting: relabelling, down-weighting, or a robust loss
that tolerates the poisons the audit misses. On the attack side, an adversary
that re-optimises the expert models under the stealth constraint would tighten
the bound our current adaptive attacker only gestures at. Finally, the
stealthy-flip pool is computable for all 90 ordered CIFAR-10 pairs in seconds,
so testing whether it predicts the trained outcome on more than four of them is
limited by victim training time alone, and is worth the compute.

\input{open_science}
\input{llm_usage}
\input{ethical_considerations}

\bibliographystyle{IEEEtran}
\bibliography{refs}
\input{appendix}

\end{document}

%% file: open_science.tex
\section*{Open Science}

We release the code and results behind this paper. The repository is
at

\begin{center}
\small\url{https://github.com/ArnabDeyKabya/SATML_2027_What_Makes_a_Label_Audit_Survive_a_Label_Only_Backdoor}
\end{center}

\smallskip
\noindent\textbf{What we release.}
All of the code. On the attack side, the FLIP pipeline and the defence-aware
attacker of Section~\ref{sec:claim2}. On the defence side, the four feature
spaces and every detector we compare, plus the removal policies
(Section~\ref{sec:audits}). The baselines are included as we ran them: SPECTRE,
spectral signatures, activation clustering, Deep $k$-NN and a loss filter. A
script computes the stealthy-flip pool for all 90 CIFAR-10 pairs.

With the code come 72 run configurations, 175 run directories, the metrics each
run produced, and the tables built from them. Every number in this paper can be
read from a committed file, and the figure scripts draw from those same files.

We also release three of the four poisoned label sets: truck$\rightarrow$deer,
which came with FLIP, and car$\rightarrow$truck and cat$\rightarrow$dog, which
we made. Those three attacks can be rerun without training experts again.

\smallskip
\noindent\textbf{What we do not release, and why.}
Nothing that holds a result. The CIFAR-10 images and the DINOv2 weights come
from their own sources, so we do not copy them. Feature caches, $k$-NN graphs
and trained checkpoints are large and easy to rebuild; the repository says how,
and Table~\ref{tab:cost} says what it costs. Raw logs and per-run score arrays
are left out for the same reason, but the metrics computed from them are in.

One gap is worth naming. The dog$\rightarrow$cat label set is not in the
repository, and that is the pair several of our sharpest results come from.
What is there is enough to check those results: the configuration that builds
the labels, the manifest of every flip it chose, and the detection and
trained-model numbers for each dog$\rightarrow$cat run we report. Rebuilding
the labels themselves needs FLIP's expert-training stage first, so this pair is
more expensive to redo from scratch than the other three. We would rather say
so than let a reader find out.

\smallskip
\noindent\textbf{Running it.}
Our runs went through a shared SLURM cluster, and the batch scripts are written
for it. They are thin wrappers around one Python entry point that the
repository documents, so the scheduler is not needed and one GPU is enough.
Checking a detection number takes a feature cache, a label set and under a
second of CPU time. Checking a trained-model number takes a training run.

\smallskip
\noindent\textbf{Scope.}
Every configuration we ship is a run we actually made, and all of them are
CIFAR-10 with a ResNet-32, the setting of Section~\ref{sec:setup}. The code
inherits FLIP's other datasets and architectures, but we did not run them and
do not ship configurations for them. Section~\ref{sec:limitations} says what
that leaves untested.

%% file: llm_usage.tex
\section*{LLM usage considerations}

We used large language models while doing this work. This section says what we
used them for, what we did not use them for, and what we did to make sure they
did not damage the paper.

\smallskip
\noindent\textbf{What we used them for.}
Three things. First, code: an LLM-based coding assistant helped us write and
refactor the analysis, plotting and table-generation scripts that turn raw run
outputs into the figures and tables you see here. Second, writing: we used LLMs
to draft internal design notes, and for editorial passes over this manuscript.
Third, as a reading aid while working through related work.

We used them because this is a small team doing a large number of experimental
runs, and the scripting around those runs is repetitive work that an assistant
does quickly. That is the whole of the motivation. Nothing about the research
question needed a language model.

\smallskip
\noindent\textbf{What we did not use them for.}
No LLM appears anywhere in the method or the experiments. The audits in this
paper run on frozen DINOv2 ViT-S/14 features and ResNet-32p victim models.
Neither is a language model, and no part of the pipeline queries one. Every
number we report comes from a training or evaluation run in the released
artifact, not from a model's output. We did not train, fine-tune or evaluate
any LLM.

\smallskip
\noindent\textbf{Accountability and correctness.}
We are responsible for everything in this paper. We have reviewed all content
that an LLM helped produce, including text, code and references, as if we had
written it ourselves. LLMs were used for editorial purposes in this manuscript,
and all outputs were inspected by the authors to ensure accuracy and
originality.

This was not a formality, and we think it is worth being concrete about why.
Early in the project our internal notes carried several claims about prior work
that the prior work does not make. The most serious was the sentence the
project was built on: that existing defences are inadequate against this
attack. Reading the original paper settled it the other way. Its authors report
that SPECTRE is effective against the attack and recommend that practitioners
adopt it. The claim was fluent, it was specific, and it was wrong.

We found it by going back to the primary source instead of trusting a summary
of it. Correcting it changed the paper for the better: reproducing SPECTRE and
pricing what it costs a defender became part of the contribution, rather than a
difficulty we had assumed away. Since then we have checked every citation in
this paper against the paper it cites, and we have read every claim we
attribute to someone else in their own words.

We apply the same rule to code and results. Generated code can be wrong in ways
that still run to completion, so we re-derive results rather than accept them.
One such audit of our own evaluation harness found that we were spending
SPECTRE's removal budget across all classes when its scores are only calibrated
within a class, which made the baseline look far weaker than it is. We retracted
those numbers and rebuilt the comparison. They are not in this paper.

\smallskip
\noindent\textbf{Transparency.}
LLMs are not integral to our methodology, so there is no LLM-dependent result
here for anyone to reproduce, and no dependence on a closed model that would
make our results irreproducible once that model changes. The research question,
the experimental design and the conclusions are ours. Where an assistant
suggested a direction, we developed it and tested it ourselves before it
entered the paper, and the artifact contains the runs that decide each claim.

The limitation LLMs did introduce is the one described above: confident,
well-formed prose about prior work that does not survive a reading of that
work. Our handling of it is verification against primary sources, applied to
every citation rather than to the ones that looked suspicious. We flag it here
because a reader deciding how much to trust our related-work section is
entitled to know that we found real errors in ours and what we did about them.

\smallskip
\noindent\textbf{Responsibility.}
The compute cost of this project is dominated by the experiments, not by the
assistant. Our runs train victim models from scratch across many
configurations, and extract DINOv2 features for the full training set; the
language-model queries alongside that work are a rounding error against it. We
chose the smallest encoder that does the job, ViT-S/14 rather than a larger
variant, and we cache the extracted features so that a sweep over audit
parameters never recomputes them. Section~\ref{sec:setup} reports the
experimental configurations and Section~\ref{sec:cost} the cost accounting.

We did not collect any data to train a model, so no question of consent or of
data-holder rights arises beyond the public benchmark datasets we use, which we
cite and use under their stated terms.

%% file: ethical_considerations.tex
\section*{Ethical Considerations}

This paper builds a defence, but it also builds an attack, and we release both.
Here is what we think the risks are and what we did about them. We follow the
Menlo report's framing.

\smallskip
\noindent\textbf{Releasing attack code.}
The attack is not ours and is not new. FLIP is published, and its code and one
of its label sets are already public~\cite{jha2023flip}. We add two things. Two
more label sets, which anyone could have made from the released FLIP code. And
the defence-aware attacker of Section~\ref{sec:claim2}, which is new: it reads
our own scoring code and picks flips our audit will miss.

We release it, and we think that is right. An adaptive attacker only its
authors can run is one nobody can check, and defences tested against fixed
attacks tend not to last. The attacker still needs the training images, the
compute to train experts, and a way to get bad labels into someone's dataset.
All of that the base attack already needed. What the release buys is that
someone other than us can attack our main claim.
Section~\ref{sec:limitations} says our adaptive numbers are an upper bound on
the defence. The release is how a reader tightens it.

\smallskip
\noindent\textbf{The risk we worry about most is false confidence.}
It is the finding of Section~\ref{sec:claim2}. A defender who picks an audit by
its detection scores can end up with a worse model than one who picks by
retraining, and the audit that looks better on paper is the one that breaks
first. Someone who reads our abstract, deploys an audit, sees a high AUPRC and
calls the data clean would be making the exact mistake this paper is about. So
we tried to make that hard. Section~\ref{sec:limitations} says where the claims
stop, and it names the pairs where the usable threshold never reaches the
recall our comparisons assume. We do not claim the defence holds against a
stronger attacker than the one we ran.

\smallskip
\noindent\textbf{Data and people.}
No human subjects, no personal data, no user study. Everything runs on
CIFAR-10~\cite{krizhevsky2009cifar}, a public benchmark of ten animal and
vehicle classes, used as published. We collected nothing and scraped nothing.
We attacked no deployed system, and there is no vulnerability to report to a
vendor: this is a property of a threat model, not a bug in a product.

\smallskip
\noindent\textbf{Compute.}
The experiments are the real cost, and Section~\ref{sec:cost} reports it stage
by stage instead of leaving it vague. Two choices kept it down. We use the
smaller DINOv2 encoder by default and the larger one only as an ablation, and
we cache features once per dataset so no sweep recomputes them. The runs sat on
a shared cluster that was running anyway.

\smallskip
\noindent\textbf{On balance.}
We think the work helps more than it hurts. Defenders get a cheaper audit and,
more usefully, a reason to distrust the metric they would have picked it by.
Attackers get little they did not have. We would change our minds if the
adaptive attacker turned out to work well beyond the setting we tested, which
is one of the things the release exists to find out.

%% file: appendix.tex
\appendices

\section{Choosing $k$ without an oracle}
\label{app:kselect}

Section~\ref{sec:limitations} says a defender has no reliable rule for choosing
$k$. This appendix shows the sweep behind that. Every number comes from a CSV
committed in the released artifact.

A defender cannot pick $k$ by AUPRC, because computing AUPRC needs the poison
mask. So we tested three rules that read only the scores, and scored each
against the oracle choice.

\begin{figure}[t]
  \centering
  \includegraphics[width=\columnwidth]{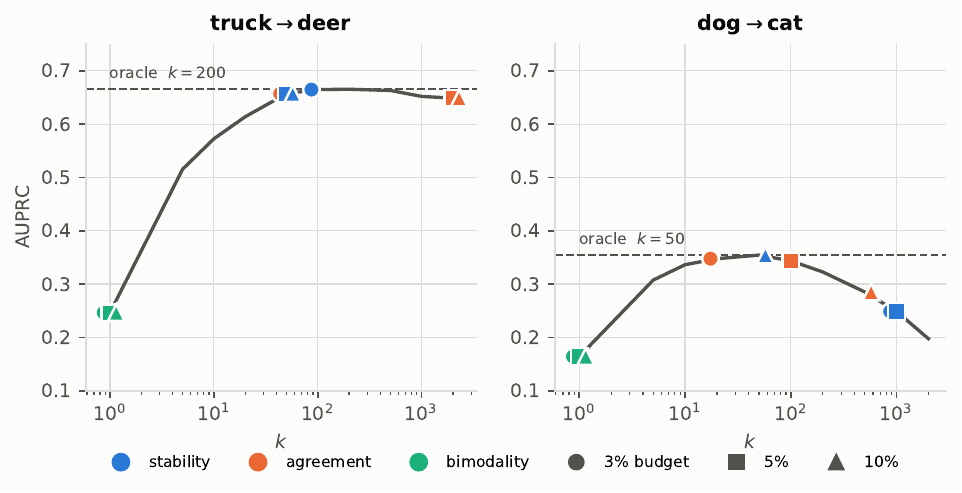}
  \caption{The AUPRC landscape over $k$ for \texttt{knn\_disagreement}, with
  what each oracle-free rule picks. The dashed line is the best AUPRC any $k$
  achieves, which a defender cannot find without the poison mask. Bimodality
  sits at $k = 1$ on both pairs and gives up most of the curve. Stability and
  agreement trade places between the pairs and between budgets. Markers are
  nudged horizontally where rules pick the same $k$; heights are exact.}
  \label{fig:kselect}
\end{figure}

The picture is not clean, and we would rather show it than summarise it.
Bimodality picks $k = 1$ on both pairs at every budget, giving up 0.42 and 0.19
AUPRC; it should not be used. The other two rules trade places. On
truck$\rightarrow$deer stability is the better rule, costing 0.0006 at the 3\%
budget and 0.0083 at 5\% and 10\%. On dog$\rightarrow$cat it is agreement that
does well at the small budgets, costing 0.0067 and 0.0108, while stability
falls to 0.1057; at the 10\% budget that reverses completely, and stability
lands exactly on the oracle while agreement gives up 0.0691.

So no rule wins everywhere, and the rule that looks best is pair-dependent and
budget-dependent. Two caveats on the rules themselves. Agreement is scored
against the prototype audit's own selection, so it is not independent of the
comparison in Section~\ref{sec:claim1} and we do not read it as evidence about
prototypes. And none of this is a defence of $k = 20$, the value every number
in this paper uses: it is not the oracle choice at either pair, and the sweep
is here to show what choosing $k$ would cost, not to justify what we chose.

\section{What a deployable threshold catches}
\label{app:thresh}

The comparison in Section~\ref{sec:spectre} cuts both detectors at 90\% recall,
which is the right way to price detectors at equal protection but is not
something a defender can dial in, because finding that cut needs the
ground-truth mask.

\begin{figure}[t]
  \centering
  \includegraphics[width=\columnwidth]{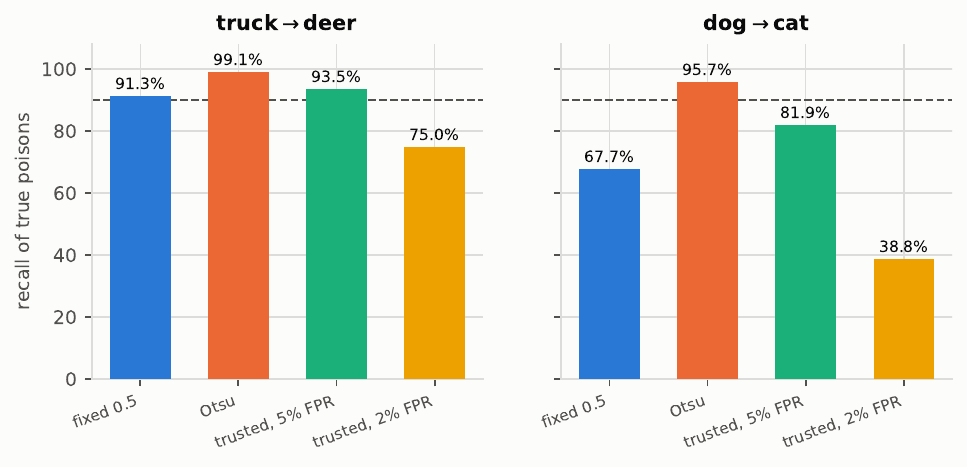}
  \caption{Recall of \texttt{knn\_disagreement} under thresholds a defender can
  set without the poison mask. The dashed line marks 90\% recall, the
  equal-protection cut used in Section~\ref{sec:spectre}; it is an oracle
  operating point, not a rule. Otsu is oracle-free and clears it on both pairs,
  but by deleting roughly twice as much; the \texttt{trusted}$n$ rules assume a
  clean subset our threat model does not grant.}
  \label{fig:thresholds}
\end{figure}

Two thresholds here need no oracle. The fixed rule catches 91.3\% on
truck$\rightarrow$deer, so the comparison point and the deployable point nearly
coincide, and 67.7\% on dog$\rightarrow$cat. Otsu, which reads only the shape
of the score histogram, catches 95.7\% on dog$\rightarrow$cat, so 90\% recall
\emph{is} reachable there without an oracle. What it costs is the budget: 5\,917
images deleted against the fixed rule's 2\,728, at an $F_1$ of 0.387 against
0.481. We did not train a victim at the Otsu cut, so we make no claim about
what that deletion buys in attack success, and a reader should treat the two
rules as different operating points rather than one dominating the other.

The \texttt{trusted}$n$ rows are a check from outside our threat model, which
grants the defender no clean subset (Section~\ref{sec:background}). We report
them because the assumption is a common one and its value should be visible: at
a 5\% false-positive rate a trusted set of 1\,000 lifts dog$\rightarrow$cat to
81.9\%, better than the fixed rule and still short of 90\%, while at 2\% it
falls to 38.8\%. The assumption helps, but it does not settle the pair.

One operating point is missing from this table. It covers
\texttt{knn\_disagreement} and \texttt{prototype\_margin}, but not
\texttt{prototype\_trimmed}, the audit we actually recommend, whose own
auto-threshold behaviour we therefore cannot report here. Its budget in
Table~\ref{tab:victim} is the one transferred from \eqref{eq:knn}
(Section~\ref{sec:audits}).

\section{The stealthy-flip pool across all 90 pairs}
\label{app:pool}

\begin{figure}[t]
  \centering
  \includegraphics[width=\columnwidth]{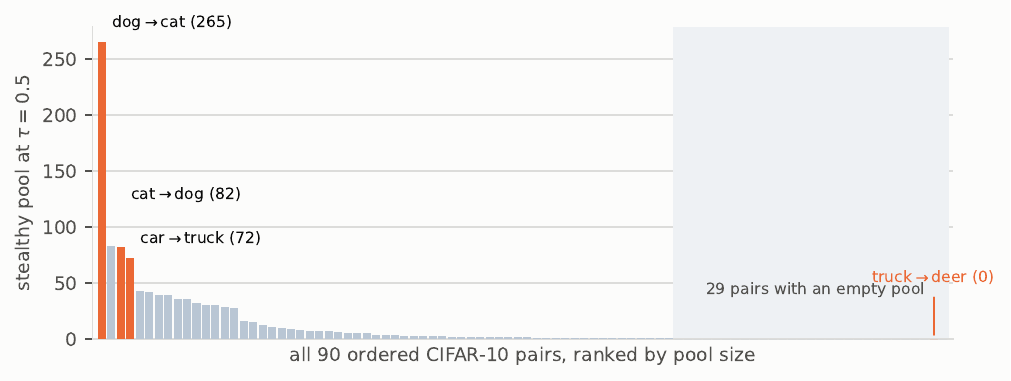}
  \caption{The stealthy-flip pool across all 90 ordered CIFAR-10 pairs, ranked.
  The four pairs we trained victims for are highlighted. One pair dominates,
  three more form a cluster behind it, and the distribution then decays into a
  long tail of pairs with no stealthy flip available at all.}
  \label{fig:pool}
\end{figure}

Figure~\ref{fig:pool} shows the whole distribution and Table~\ref{tab:pool}
lists the ten pairs with the most room for a stealthy attack.

What the pool needs is the encoder and the \emph{clean} labels
(Section~\ref{sec:claim3}), which is worth being exact about, because it
decides what the quantity is for. A defender does not have clean labels; if
they did they would not need an audit. So the pool is not a screening tool a
practitioner can run on a dataset they suspect. It is an analysis tool: it lets
\emph{us} say, before running any attack, which pairs give an attacker room,
and it is computed here on all 90 pairs in seconds with no training. Whether
the same zero-versus-non-zero split survives when the pool is computed on
poisoned labels instead is a cheap check and a fair thing to ask; we have not
run it.

Two things in this table bound what Section~\ref{sec:claim3} can claim. The top
of the ranking is a cluster: dog$\rightarrow$cat has 265, then a gap to 83, 82
and 72, and the fifth pair has 43. We trained victims on three of those four;
the one we did not, truck$\rightarrow$car, is the inverse of
car$\rightarrow$truck and the untested direction among the hardest pairs. And
the pool does not order pairs within the non-empty group at all, so nothing
here supports reading it as a severity score.

\section{The defence-aware attacker}
\label{app:adaptive}

Section~\ref{sec:claim2} rests on this attacker, so we state it precisely.

The attacker knows the encoder, $k$, the scoring rule and the threshold, and
can evaluate the defender's score for any candidate flip before committing to
it. Writing $\mathrm{margin}(i)$ for the FLIP objective that the
defence-unaware attacker maximises, and $\mathrm{score}(i, \ell)$ for the
defender's score on sample $i$ carrying label $\ell$, it solves
\begin{equation*}
\begin{aligned}
  \max_{S} \quad & \textstyle\sum_{i \in S} \mathrm{margin}(i) \\[2pt]
  \text{s.t.} \quad & |S| = N, \\[2pt]
                    & \mathrm{score}\big(i, \ell_{\text{flip}}(i)\big) \le \tau
                      \quad \forall i \in S,
\end{aligned}
\end{equation*}
with $N = 1500$ and $\tau$ the stealth budget. Because the constraint is
per-sample and the objective is a sum, the solution is exact and needs no
search: mask every ineligible sample and take the $N$ largest margins among the
rest. At $\tau = 1$ the constraint is vacuous and the selection reproduces the
defence-unaware attacker exactly, which the released code asserts on every run
rather than assuming.

Two details matter for reading the numbers. First, the defender scores a sample
against the labels its neighbours \emph{carry}, which after the attack includes
other poisons, so a cluster of co-located flips lowers its own scores. We let
the attacker have that: the selection is re-run to a fixed point over the
poisoned assignment (three iterations) rather than scored once against clean
labels. Second, the attacker chooses \emph{which} samples to flip but not what
to flip them to; the target label stays as FLIP's optimisation assigned it.
Re-optimising the label assignment jointly with the stealth constraint is a
strictly stronger attack, so every adaptive number in this paper bounds the
defence from above, and we sweep $\tau$ from 1.0 down to 0.1 to trace that
bound rather than report a single point.

\input{appendix_tables}

%% file: appendix_tables.tex
\begin{table}[t]
  \centering
  \caption{Choosing $k$ without an oracle, for \texttt{knn\_disagreement} on
  frozen DINOv2 features. For each pair and audit budget $B$ we give the $k$
  that maximises AUPRC (the oracle a defender cannot consult) and what each
  oracle-free rule picks instead, with the AUPRC it gives up. AUPRC ranks the
  whole score vector and so does not depend on $B$; the oracle column is
  therefore constant down a pair, and only the rules move with the budget.
  \emph{Stability} takes the smallest $k$ whose top-$B$ selected set has
  Jaccard overlap at least 0.85 with the top-$B$ set at $2k$, which is why it
  is undefined where $2k$ is off the grid. \emph{Bimodality} maximises Otsu
  separation of the score histogram. \emph{Agreement} maximises top-$B$ overlap
  with \texttt{prototype\_trimmed}, so it is not independent of the prototype
  audit and we do not read it as evidence about prototypes.}
  \label{tab:kselect}
  \small
  \setlength{\tabcolsep}{4pt}
  \begin{tabular}{llcllc}
    \toprule
    pair & budget & oracle $k$ & rule & picks $k$ & AUPRC gap \\
    \midrule
    truck$\rightarrow$deer & 3\% & 200 & \texttt{stability} & 100 & $+0.0006$ \\
     &  &  & \texttt{bimodality} & 1 & $+0.4190$ \\
     &  &  & \texttt{agreement} & 50 & $+0.0083$ \\
    \addlinespace
     & 5\%, 10\% & 200 & \texttt{stability} & 50 & $+0.0083$ \\
     &  &  & \texttt{bimodality} & 1 & $+0.4190$ \\
     &  &  & \texttt{agreement} & 2000 & $+0.0167$ \\
    \addlinespace
    dog$\rightarrow$cat & 3\% & 50 & \texttt{stability} & 1000 & $+0.1057$ \\
     &  &  & \texttt{bimodality} & 1 & $+0.1901$ \\
     &  &  & \texttt{agreement} & 20 & $+0.0067$ \\
    \addlinespace
     & 5\% & 50 & \texttt{stability} & 1000 & $+0.1057$ \\
     &  &  & \texttt{bimodality} & 1 & $+0.1901$ \\
     &  &  & \texttt{agreement} & 100 & $+0.0108$ \\
    \addlinespace
     & 10\% & 50 & \texttt{stability} & 50 & $+0.0000$ \\
     &  &  & \texttt{bimodality} & 1 & $+0.1901$ \\
     &  &  & \texttt{agreement} & 500 & $+0.0691$ \\
    \bottomrule
  \end{tabular}
\end{table}

\begin{table*}[t]
  \centering
  \caption{What a deployable threshold actually catches. \texttt{fixed\_0.5}
  and \texttt{otsu} are both oracle-free: the first is a fixed cut, the second
  thresholds the score histogram by maximising between-class variance and uses
  no labels or clean data of any kind. Only the \texttt{trusted}$n$ rules
  assume a clean subset, of size $n$, at the stated false-positive rate; our
  threat model does not grant one (Section~\ref{sec:background}), so those rows
  are a check from outside it. Recall is of the true poison mask. The 90\%
  recall used for the equal-protection comparison in Section~\ref{sec:spectre}
  is an oracle operating point, not one of these.}
  \label{tab:thresholds}
  \small
  \setlength{\tabcolsep}{3pt}
  \begin{tabular}{lllrrr}
    \toprule
    pair & detector & rule & flagged & recall & $F_1$ \\
    \midrule
    truck$\rightarrow$deer & \texttt{knn\_disag.} & \texttt{fixed\_0.5} & 2981 & 91.3\% & 0.611 \\
     &  & \texttt{otsu} & 5746 & 99.1\% & 0.410 \\
     &  & \texttt{trusted1000\_fpr0.05} & 3367 & 93.5\% & 0.576 \\
     &  & \texttt{trusted1000\_fpr0.02} & 1992 & 75.0\% & 0.644 \\
    \addlinespace
     & \texttt{proto\_margin} & \texttt{fixed\_0.5} & 61 & 4.1\% & 0.078 \\
     &  & \texttt{otsu} & 1405 & 88.3\% & 0.912 \\
     &  & \texttt{trusted1000\_fpr0.05} & 3661 & 99.3\% & 0.577 \\
     &  & \texttt{trusted1000\_fpr0.02} & 2457 & 98.5\% & 0.747 \\
    \addlinespace
    dog$\rightarrow$cat & \texttt{knn\_disag.} & \texttt{fixed\_0.5} & 2728 & 67.7\% & 0.481 \\
     &  & \texttt{otsu} & 5917 & 95.7\% & 0.387 \\
     &  & \texttt{trusted1000\_fpr0.05} & 3739 & 81.9\% & 0.469 \\
     &  & \texttt{trusted1000\_fpr0.02} & 1492 & 38.8\% & 0.389 \\
    \addlinespace
     & \texttt{proto\_margin} & \texttt{fixed\_0.5} & 3 & 0.2\% & 0.004 \\
     &  & \texttt{otsu} & 1630 & 70.8\% & 0.679 \\
     &  & \texttt{trusted1000\_fpr0.05} & 3607 & 92.2\% & 0.542 \\
     &  & \texttt{trusted1000\_fpr0.02} & 1838 & 74.4\% & 0.669 \\
    \bottomrule
  \end{tabular}
\end{table*}

\begin{table*}[t]
  \centering
  \caption{Stealthy-flip pool for all 90 ordered CIFAR-10 pairs, computed on frozen DINOv2 features in seconds and with no training. The pool counts source-class images whose neighbourhood already leans to the target strongly enough that a flip survives the audit at $\tau$. The ten largest are shown; 29 of the 90 pairs have an empty pool at $\tau = 0.5$. Pairs we trained victims for are marked $\star$.}
  \label{tab:pool}
  \small
  \begin{tabular}{lrrr}
    \toprule
    pair & $\tau = 0.5$ & $\tau = 0.7$ & $\tau = 0.9$ \\
    \midrule
    dog$\rightarrow$cat$\star$ & 265 & 527 & 1164 \\
    truck$\rightarrow$car & 83 & 164 & 425 \\
    cat$\rightarrow$dog$\star$ & 82 & 314 & 1223 \\
    car$\rightarrow$truck$\star$ & 72 & 143 & 402 \\
    bird$\rightarrow$plane & 43 & 68 & 165 \\
    deer$\rightarrow$frog & 42 & 79 & 150 \\
    bird$\rightarrow$frog & 39 & 87 & 213 \\
    cat$\rightarrow$frog & 39 & 104 & 310 \\
    plane$\rightarrow$ship & 36 & 61 & 176 \\
    bird$\rightarrow$deer & 36 & 88 & 267 \\
    \addlinespace
    truck$\rightarrow$deer$\star$ & 0 & 0 & 8 \\
    \bottomrule
  \end{tabular}
\end{table*}